\documentclass[aos,preprint]{imsart}
\RequirePackage{amsthm,amsmath,amsfonts,amssymb}
\usepackage{mathtools}
\RequirePackage{natbib}
\usepackage{mathrsfs}
\usepackage{amsfonts}
\usepackage{makeidx}         
\RequirePackage{graphicx}        
\usepackage{multicol}        
\usepackage[bottom]{footmisc}
\usepackage{epsfig}
\usepackage{bm}
\usepackage{enumitem}
\usepackage{epstopdf}
\usepackage{multirow}
\usepackage{booktabs}
\usepackage{mathtools}
\usepackage{wrapfig}
\usepackage[pagewise, mathlines]{lineno}
\usepackage{caption}
\usepackage{subcaption}
\usepackage{lscape}
\usepackage[colorinlistoftodos]{todonotes}
\usepackage{float}
\usepackage{titlesec}
\usepackage{verbatim}
\usepackage{rotating}
\usepackage{siunitx}
\usepackage{booktabs}
\usepackage[allcolors=blue,colorlinks=true]{hyperref}

\theoremstyle{plain}

\newtheorem{theorem}{Theorem}
\newtheorem{lemma}{Lemma}
\newtheorem{proposition}{Proposition}
\newtheorem{corollary}{Corollary}
\newtheorem{condition}{condition}
\theoremstyle{remark}
\newtheorem{definition}[theorem]{Definition}

\newlength{\tempheight}
\newlength{\tempwidth}

\newcommand{\rowname}[1]
{\rotatebox{90}{\makebox[\tempheight][c]{\textbf{#1}}}}

\newcommand{\columnname}[1]
{\makebox[\tempwidth][c]{\textbf{#1}}}

\mathtoolsset{showonlyrefs}
\startlocaldefs
\theoremstyle{remark}
\newtheorem{algorithm}{Algorithm}
\newtheorem{remark}{Remark}

\endlocaldefs

\begin{document}

\begin{frontmatter}
\title{Efficiency Requires Adaptation}
\runtitle{Efficiency Requires Adaptation}

\begin{aug}
\author{\fnms{Adam} \snm{Lane}\ead[label=e1]{adam.lane@cchmc.org}},
\address{University of Cincinnati}
\end{aug}

\begin{abstract}
The majority of historical designs are \emph{a priori} in nature, where \emph{a priori} indicates a design can be specified in advance of the experiment. The conventional wisdom is that the set of \emph{a priori} designs is sufficient to produce efficient experiments. This work challenges this convention and finds that efficiency requires data dependent strategies. Specifically, in the context of a sequential experiment, where observations are accrued in a series of runs, an adaptive design is proposed that is guaranteed to be more efficient than any corresponding \emph{a priori} design.   
\end{abstract}



\end{frontmatter}

\section{Introduction} \label{sec:Intro}

The Cram\'{e}r Rao lower bound (CRLB) provides an explicit bound on the achievable precision of any parameter estimate obtained following an experiment. The CRLB(${\xi}$) is a function of the design, $\xi=[\boldsymbol{x}(1), \ldots, \boldsymbol{x}(n)]^{T}$, where $\boldsymbol{x}(j)$ represents the support point of observation $j (=1,\ldots,n)$. A critical challenge for an experimenter is to select a design that is efficient with respect to the CRLB. 

Different experimental settings require different design strategies. Factorial designs are commonly used in experiments with many treatment factors. Optimal designs minimize the CRLB with respect to a specific convex optimality criterion. Minimax designs are robust against worst case scenarios. While, each of these designs are valuable in the appropriate context, they also, often, satisfy a broader definition of efficiency. Specifically, a design $\xi_{1}$ \textbf{is} \emph{efficient} if there \textbf{does not exist} a design, $\xi_{2}$, such that CRLB(${\xi_{1}}$) > CRLB(${\xi_{2}}$) in Loewner order. For two symmetric matrices $\boldsymbol{A}$ and $\boldsymbol{B}$, $\boldsymbol{A}>\boldsymbol{B}$ in Loewner order if $\boldsymbol{A}-\boldsymbol{B}$ is positive definite. 

While the CRLB is the most popular lower bound there exist tighter alternatives [\cite{Bhat:OnSo:1948,VanT:Dete:1968,Bobr:Maye:Zaka:Some:1987}]. A design is more robustly efficient if it satisfies an analogous definition with respect to a tighter lower bound. In Section \ref{sec:LB} such a lower bound is introduced. Efficiency, as considered in this work, is formally defined with respect to this alternative lower bound [Section \ref{sec:RSD}].

Historically, the design literature has primarily focused on \emph{a priori} designs - designs that can be specified in advance of the experiment. The set of \emph{a priori} designs includes deterministic designs [eg. factorial, optimal and minimax designs] and designs selected via data independent random processes [e.g. \cite{Wu:OnTh:1981,Li:Mini:1983,Wait:Wood:Mini:2021}].

Conventional wisdom suggests that the set of \emph{a priori} designs is sufficient to produce efficient experiments. The main result of this work finds this convention is incorrect; efficiency requires data dependent strategies. This apparent paradox is proven by developing an adaptive design that when applied to any \emph{a priori} design reduces the lower bound [proposed in Section \ref{sec:LB}] in Loewner order. Adaptive designs can be implemented in experiments that are observed in a series of sequential runs and are characterized by the use of the data accrued from preceding runs to determine the design of the current run. 

\subsection{Robust Adaptive Designs}

Adaptive designs have been employed in experiments where the CRLB depends on the unknown parameters. In such cases \emph{a priori} designs are only efficient locally at the true value of the unknown parameters. As an illustration, consider an experiment with two treatments, denoted 1 and 2, with means $\eta_{1}$ and $\eta_{2}$ and variances $\sigma_{1}^{2}$ and $\sigma_{2}^{2}$, respectively. Neyman allocation states that an \emph{a priori} design that allocates $\rho = \sigma_{1}/(\sigma_{1} + \sigma_{2})$ proportion of the observations to treatment 1, and the remaining observations to treatment 2, is efficient for estimating the difference in the means. If $\rho$ is unknown then one could guess its value and use a design that is efficient with respect to this guess. This strategy can be inefficient if the guess is far from reality. An adaptive design that assigns observation $(j+1)$ to treatment $1$ with probability $\hat{\rho}(j)$ is a robust alternative to guessing, where $\hat{\rho}(j)$ is an estimate of $\rho$ based on the first $j$ observations [\cite{Hu:Rose:Opti:2003}]. Designs that adaptively estimate the parameters to produce robust experiments have received significant attention [\cite{Box:Hunt:Sequ:1965,Lai:Robb:Adap:1979,Fedo:Leon:Opti:2013,Lane:Yao:Flou:Info:2014,Kim:Flou:Opti:2014}].

While the described adaptive design is more robust than guessing, it is not more efficient than the \emph{a priori} design that allocates $\rho$ observations to treatment 1. Even though this ideal design is unknown, and cannot be used in practice, it is included in the set of \emph{a priori} designs. In contrast to robust adaptive designs, the adaptive design framework developed in this work incorporates ancillary statistics into the design and is more efficient than any \emph{a priori} design. Ancillary statistics, by definition, contain no information about the model parameters; however, certain ancillary statistics impact the quality of the parameter estimates [\cite{Fish:TwoN:1934,Cox:Some:1958,Efro:Hink:Asse:1978}]. \cite{Fish:TwoN:1934}, \cite{Basu:Reco:1969} and \cite{Ghos:Reid:Fras:Anci:2010} refer to such ancillary statistics as a reference set or \emph{relevant subset} - a subset of the sample space on which inference should be restricted. Post-data conditional inference on relevant subsets has been thoroughly investigated [\cite{Fish:TwoN:1934,Cox:Some:1958,Basu:Reco:1969,McCu:Cond:1992,Efro:Hink:Asse:1978,Ghos:Reid:Fras:Anci:2010}]. An illustrative example is provided in Section \ref{sec:LB}.

\subsection{Summary of the Contributions}
Recently, \cite{lane2020efficiency,Lane:Adap:2019,lane2021optimal} proposed designs that incorporate the relevant subset, refered to as relevant subset designs (RSDs), into an optimal design framework. The current work significantly extends the literature on RSD, and more broadly experimental design, in several key ways. 

In Section \ref{sec:LB} a lower bound is derived from the relevant subset. This alternative bound is tighter than the CRLB and provides a metric to evaluate the efficiency of RSDs. In Section \ref{sec:RSD} a novel RSD is developed with the expressed intent of reducing the proposed lower bound. This approach places limited restrictions on the experimental design. The setting could correspond to a factorial, optimal, minimax, randomized design, etc. Additionally, there is limited restriction on the method used to estimate $\boldsymbol{\theta}$. This represents a significant extension of the existing literature on RSDs which has currently only considered the maximum likelihood estimate (MLE) in the context of optimal design.  

In Section \ref{sec:Main} the theoretical benefits of the proposed RSD are presented. The primary result is that given any \emph{a priori} design the proposed RSD increases the efficiency in Loewner order. This is a significant generalization of the theoretical benefits associated with existing RSDs which have only been shown to be efficient, primarily through heuristics and simulations, with respect to a single convex optimality criterion. It should be noted that this is all done in the context of a general nonlinear model framework. In Section \ref{sec:Examples} a simulation study is conducted and it is found that the proposed RSD is more efficient than pertinent alternatives in every case considered.  

\section{Lower Bounds on Variance} \label{sec:LB}
One of the most common experimental objectives is to characterize the relationship between a sequence of univariate responses, $\boldsymbol{y} = [y(1),\ldots,y(n)]^{T}$, and a set of corresponding explanatory variables $\boldsymbol{x}(1),\ldots,\boldsymbol{x}(n)$. In this work the set of explanatory variables, denoted $\xi = [\boldsymbol{x}(1),\ldots,\boldsymbol{x}(n)]$, is assigned by the experimenter and is referred to as the design where $\boldsymbol{x}(j)$ is a $s\times 1$ dimensional vector for $j = 1,\ldots,n$. This relationship is often assessed using an additive error regression model of the form
\begin{align} \label{eq:model}
y(j) = \eta[\boldsymbol{x}(j),\boldsymbol{\theta}] + \varepsilon(j), 
\end{align}
where $\eta(\boldsymbol{x},\boldsymbol{\theta})$ is a, potentially nonlinear, function that relates $\boldsymbol{x}$ to $y$ through the $p$-dimensional parameter of interest $\boldsymbol{\theta}$ subject to an error term $\varepsilon$.

Let $\boldsymbol{X} = (\boldsymbol{x}_{1},\ldots,\boldsymbol{x}_{d})^{T}$ denote the unique levels of the explanatory variables used in an experiment and $\boldsymbol{w} = (w_{i},\ldots,w_{d})^{T}$ represents the corresponding allocation weights, where $w_{i}=n_{i}/n$, $n_{i} = \sum_{j}t_{i}(j)$, $t_{i}(j)$ is an indicator variable such that $t_{i}(j) = 1$ if $\boldsymbol{x}(j)=\boldsymbol{x}_{i}$ and 0 otherwise and $n=\sum_{i}n_{i}$. 
\begin{definition} \label{def:FixedDesign}
For a given design space, $\mathscr{X}$, a design $\xi$ is an $n$-point \emph{a priori} deterministic design if (1) $\boldsymbol{x}_{i}\in\mathscr{X}$, where $\mathscr{X}$ is a compact subset of $\mathbb{R}^{s}$; (2) $0\le w_{i} \le 1$, $\sum w_{i}=1$ and $nw_{i}$ are integers for all $i=1,\ldots,d<\infty$; and (3) $n$ is a pre-specified constant. 
\end{definition}
Let the set $\Xi_{n}$ represent the collection of all $n$-point \emph{a priori} deterministic designs. As outlined in the introduction the set of \emph{a priori} designs includes both deterministic and data independent random strategies. For the latter, the design used in the experiment, $\xi$, is an observed realization from a probability measure $\pi$ with corresponding probability space $(\mathscr{X},\Xi_{n},\pi)$. The probability measure $\pi$ is an \emph{a priori} random design if $\pi(\xi|\boldsymbol{\mathcal{Y}}) = \pi(\xi)$, where $\boldsymbol{\mathcal{Y}} = [\mathcal{Y}(1),\ldots,\mathcal{Y}(n)]^{T}$ is the random response vector. Calligraphic capital letters indicate random variables; whereas, roman letters correspond to observed realizations. This is a deviation from the standard practice of using capital roman letters to indicate random variables. This notation was implemented to distinguish random variables from observed matrices, which are indicated with bold capital roman letters, e.g. $\boldsymbol{X}$. The term \emph{a priori} references the fact that, due to their independence of the responses, they can be specified in advance of the experiment. The set of deterministic designs is a subset of the randomized designs; i.e. any deterministic design $\xi\in\Xi_{n}$ can be obtained by setting $\pi$ equal to $\delta_{\xi}$, where $\delta_{\xi}$ is a measure that assigns unit mass to $\xi$. 

\subsection{Cram\'{e}r-Rao Lower Bound}

A critical question for an experimenter is to determine a design that produces efficient estimates of $\boldsymbol{\theta}$. Due to its connection to the CRLB the Fisher information in the sample is an important quantity in evaluating this objective. 

In the context of model \eqref{eq:model} the Fisher information in the sample on $\boldsymbol{\theta}$ can be constructed from the elemental information. \cite{Atki:Fedo:Herz:Elem:2014} define the \emph{elemental information} as the information from a single observation associated with a distribution in its standard form, as opposed to with respect to the parameterization induced by the inclusion of explanatory variables. Let $L(\eta|y) = f(y|\eta)$, $l_{y} = \log L(\eta|y)$, $\dot{l}_{y} = (\partial/\partial\eta) \log L(\eta|y)$, $\ddot{l}_{y} = (\partial/\partial\eta)  \dot{l}_{y}$ and $i_{y} = -\ddot{l}_{y}$, where $f(y|\eta)$ is the probability density function (PDF) of $y$. Consider the following condition.

\begin{condition} \label{cond:Regular}
(1) $\boldsymbol{\varepsilon} = [\varepsilon(1),\ldots,\varepsilon(n)]^{T}$ is a sequence of independent and identically distributed random variables; (2) $\boldsymbol{\theta}\in\Theta$, where $\Theta$ is an open set in $\mathbb{R}^{p}$ (3) $(\partial/\partial\boldsymbol{\theta})\eta(\boldsymbol{x},\theta)$ and $(\partial^{2}/\partial\boldsymbol{\theta}^{2})\eta(\boldsymbol{x},\theta)$ exist for all $\boldsymbol{x}\in\mathscr{X}$; (4) $(\partial/\partial\eta)f(y|\eta)$ and $(\partial^{2}/\partial\eta^{2})f(y|\eta)$ exist; and (5) $(\partial/\partial\eta)$ can be passed twice under the integral sign in $\int f(y|\eta)dy$ and once in $\int \tilde{\boldsymbol{\theta}} f(y|\eta)dy$. 
\end{condition}

Under Condition \ref{cond:Regular} the elemental information $\mu = {\rm{E}}[i_{\mathcal{Y}}]$ is non-negative. For clarity of exposition $\mu$ is a shared term for all observations. Cases where this is violated can often be transformed into the current setting. For example, suppose the elemental information is $\mu s(\boldsymbol{x})$, where $s(\boldsymbol{x})$ is a known function of $\boldsymbol{x}$; multiplying $y$ and $\eta(\boldsymbol{x},\boldsymbol{\theta})$ by $s^{1/2}(\boldsymbol{x})$ reduces to the current setting. 

Let $\eta_{i} = \eta(\boldsymbol{x}_{i},\boldsymbol{\theta})$, $L(\eta_{i}|\boldsymbol{y},\boldsymbol{t}_{i}) = \prod_{j=1}^{n} L[\eta_{i}|y(j)]^{t_{i}(j)}$, $l_{\boldsymbol{y},\boldsymbol{t}_{i}} = \sum_{j=1}^{n} t_{i}(j) l_{y(j)}$ and $\dot{l}_{\boldsymbol{y},\boldsymbol{t}_{i}} = \sum_{j=1}^{n} t_{i}(j) \dot{l}_{y(j)}$. The observed information on $\eta_{i}$ is $i_{\boldsymbol{y},\boldsymbol{t}_{i}} = \sum_{j=1}^{n} t_{i}(j) i_{y(j)}$, where $\boldsymbol{t}_{i} = [t_{i}(1),\ldots,t_{i}(n)]^{T}$. Note, observations with $t_{i}(j)=0$ do not contribute to $i_{\boldsymbol{y},\boldsymbol{t}_{i}}$. The Fisher information in the sample, for a given the design $\xi$, on $\eta_{i}$ is ${\rm{E}}[i_{\boldsymbol{\mathcal{Y}},\boldsymbol{\mathcal{T}}_{\negmedspace i}}\mid\xi] = nw_{i}\mu$. The preceding expectation implicitly assumes $\xi$ is an \emph{a priori} deterministic design; the extension of the results in this work to random design strategies is considered in Sections \ref{subsec:Random} and \ref{subsec:MainRandom}.

Switching to the Fisher information in the sample on $\boldsymbol{\theta}$, denote the likelihood of $\boldsymbol{\theta}$ as $L(\boldsymbol{\theta}|\boldsymbol{y},\boldsymbol{T}) = \prod_{i=1}^{d} L(\eta_{i}|\boldsymbol{y},\boldsymbol{t}_{i})$. The observed information on $\boldsymbol{\theta}$ is
\begin{align}
\boldsymbol{I}_{\boldsymbol{y},\boldsymbol{T}}(\boldsymbol{\theta}) = -\frac{\partial^{2}}{\partial \boldsymbol{\theta}^{2}}\log L(\boldsymbol{\theta}|\boldsymbol{y},\boldsymbol{T}) =  \sum_{j=1}^{n} \left[ i_{\boldsymbol{y},\boldsymbol{t}_{i}} \frac{\partial \eta(\boldsymbol{x}_{i},\boldsymbol{\theta})}{\partial\boldsymbol{\theta}} \frac{\partial \eta(\boldsymbol{x}_{i},\boldsymbol{\theta})}{\partial\boldsymbol{\theta}^{T}} - \dot{l}_{\boldsymbol{y},\boldsymbol{t}_{i}} \frac{\partial^{2} \eta(\boldsymbol{x}_{i},\boldsymbol{\theta})}{\partial\boldsymbol{\theta}^{2}} \right],
\end{align}
where $\boldsymbol{T} = (\boldsymbol{t}_{1},\ldots,\boldsymbol{t}_{d})^{T}$ is observed allocation matrix. Under Condition \ref{cond:Regular} the Fisher information in the sample on $\boldsymbol{\theta}$ is 
\begin{align}
    \boldsymbol{F}_{\xi}(\boldsymbol{\theta}) = {\rm{E}}\left[\boldsymbol{I}_{\boldsymbol{\mathcal{Y}},\boldsymbol{\mathcal{T}}}(\boldsymbol{\theta}) \mid \xi \right] = n\mu\sum_{i=1}^{d}w_{i}\frac{\partial \eta(\boldsymbol{x}_{i},\boldsymbol{\theta})}{\partial\boldsymbol{\theta}} \frac{\partial \eta(\boldsymbol{x}_{i},\boldsymbol{\theta})}{\partial\boldsymbol{\theta}^{T}},
\end{align}
where, in the present notation $\boldsymbol{\mathcal{Y}}$ and $\boldsymbol{\mathcal{T}}$ are the random response vector and allocation matrix, respectively. The dependence of $\boldsymbol{F}_{\xi}$ on $\boldsymbol{\theta}$ will be omitted when the meaning is clear. Note that the design, $\xi$, completely determines $\boldsymbol{\mathcal{T}}$ and that it is assumed that the design $\xi$ results in a positive definite $\boldsymbol{F}_{\xi}$.

The most commonly used lower bound on the variance of an estimate, say $\tilde{\boldsymbol{\theta}}$, with finite expectation $\boldsymbol{b}(\boldsymbol{\theta}) = {\rm{E}}[\tilde{\boldsymbol{\theta}}|\xi]$ is the Cram\'{e}r-Rao Lower Bound (CRLB) which states, under Condition \ref{cond:Regular}, that
\begin{align} \label{eq:BiasCRLB}
    {\rm{Var}}[\tilde{\boldsymbol{\theta}}|\xi] \ge \frac{\partial \boldsymbol{b}(\boldsymbol{\theta})}{\partial\boldsymbol{\theta}}\boldsymbol{F}_{\xi}^{-1}\frac{\partial \boldsymbol{b}(\boldsymbol{\theta})}{\partial\boldsymbol{\theta}^{T}}.
\end{align}
In many settings asymptotically unbiased estimators are of interest. For example, the MLE is unbiased to the order $O(n^{-1})$ which implies that $(\partial/\partial\boldsymbol{\theta})\boldsymbol{b}(\boldsymbol{\theta}) = \boldsymbol{I}_{p}\{1 + O(n^{-1})\}$, where $\boldsymbol{I}_{p}$ is the $p\times p$ identity matrix. Bias correction methods can be implemented to further reduce the bias if desired [\cite{Hall:Mart:OnBo:1988,Firt:Bias:1993,Pron:Pazm:Bias:1994,Kosm:Firt:Bias:2009,Cord:Crib:AnIn:2014}]. Since, the bias has a negligible impact, as $n$ gets large, a common approach in the design of experiments is to consider the unbiased version of the CRLB which states that if ${\rm{E}}[\tilde{\boldsymbol{\theta}}|\xi] = \boldsymbol{\theta}$ then 
\begin{align} \label{eq:CRLB}
    {\rm{Var}}[\tilde{\boldsymbol{\theta}}|\xi] \ge \boldsymbol{F}_{\xi}^{-1}.
\end{align}
Unless explicitly stated otherwise all future references to the CRLB references \eqref{eq:CRLB}. The CRLB has been a significant motivator for designing experiments that are efficient with respect to $\boldsymbol{F}_{\xi}^{-1}$. The conventional wisdom being that reducing the lower bound improves estimation and inference following an experiment. 

\subsection{A Relevant Subset Lower Bound}
This section develops an alternative lower bound based on the relevant subset. The relevant subset is embedded in the definition of sufficiency. Suppose $(\tilde{\boldsymbol{\theta}},\boldsymbol{A})$ is a sufficient statistic, where $\tilde{\boldsymbol{\theta}}$ is an estimate of $\boldsymbol{\theta}$ and $\boldsymbol{A}$ is an ancillary random variable. In such cases $\tilde{\boldsymbol{\theta}}$ is not sufficient and inference based on  $\tilde{\boldsymbol{\theta}}$ alone results in a loss of information. \cite{Fish:TwoN:1934} argued that this information is recovered by conditioning on the \emph{relevant subset}, the subset of the sample space where $\boldsymbol{\mathcal{A}} = \boldsymbol{A}$, and by extension that the Fisher information in the relevant subset, derived below, is more meaningful than the Fisher information in the sample. Recall, in the current notation $\boldsymbol{\mathcal{A}}$ is a random ancillary variable and $\boldsymbol{A}$ is its observed realization.

Analogous to the derivation of $\boldsymbol{F}_{\xi}$, the Fisher information in the relevant subset on $\boldsymbol{\theta}$ is constructed from the Fisher information in the relevant subset on $\eta_{i}$. Let $f(\boldsymbol{y}_{i}|\boldsymbol{a}_{i},\eta_{i})$ be the conditional probability density of $\boldsymbol{y}_{i} = \{y(j):t_{i}(j)=1, j=1,\ldots,n\}$, where $\boldsymbol{a}_{i}$ is an ancillary random variable constructed from the data corresponding to the support point $\boldsymbol{x}_{i}$. For this section it is required to replace Condition \ref{cond:Regular} with the more restrictive condition stated below. 

\begin{condition} \label{cond:RegularExt}
Condition \ref{cond:Regular} (1-4); (5') $(\partial/\partial\eta_{i})$ can be passed twice under the integral sign in $\int f(\boldsymbol{y}_{i}|\boldsymbol{a}_{i},\eta_{i})d\boldsymbol{y}_{i}|\boldsymbol{a}_{i}$ and once in $\int \tilde{\boldsymbol{\theta}} f(\boldsymbol{y}_{i}|\boldsymbol{a}_{i},\eta_{i})d\boldsymbol{y}_{i}|\boldsymbol{a}_{i}$  for all $\boldsymbol{a}_{i}\in \mathscr{S}_{i}$, where $\mathscr{S}_{i}$ is the support of $\boldsymbol{a}_{i}$, $i=1,\dots,d$; (6) $\boldsymbol{b}_{\boldsymbol{A}} (\boldsymbol{\theta}) = {\rm{E}}[\tilde{\boldsymbol{\theta}}|\boldsymbol{\mathcal{A}} = \boldsymbol{A},\xi] = \boldsymbol{\theta}$ for all $\boldsymbol{A}\in \mathscr{S}$, where $\mathscr{S}$ is the support of $\boldsymbol{A} = (\boldsymbol{a}_{1},\ldots,\boldsymbol{a}_{d})^{T}$; and (7) $\boldsymbol{F}_{\xi}$ is positive definite.
\end{condition}

The Fisher information in the relevant subset on $\eta_{i}$ is $h_{\boldsymbol{\mathcal{A}}_{i}} = {\rm{E}}[i_{\boldsymbol{\mathcal{Y}},\boldsymbol{\mathcal{T}}_{\negmedspace i}}|\boldsymbol{\mathcal{A}}_{i},\xi]$ and the Fisher information in the relevant subset on $\boldsymbol{\theta}$ is 
\begin{align} \label{eq:RelInfo}
    \boldsymbol{H}_{\boldsymbol{\mathcal{A}}}(\boldsymbol{\theta})  = {\rm{E}}\left[\left. \boldsymbol{I}_{\boldsymbol{\mathcal{Y}},\boldsymbol{\mathcal{T}}}(\boldsymbol{\theta}) \right| \boldsymbol{\mathcal{A}},\xi\right] = \sum_{i=1}^{d}h_{\boldsymbol{\mathcal{A}}_{i}}\frac{\partial \eta(\boldsymbol{x}_{i},\boldsymbol{\theta})}{\partial\boldsymbol{\theta}} \frac{\partial \eta(\boldsymbol{x}_{i},\boldsymbol{\theta})}{\partial\boldsymbol{\theta}^{T}},
\end{align}
where $\boldsymbol{\mathcal{A}} = (\boldsymbol{\mathcal{A}}_{1},\ldots,\boldsymbol{\mathcal{A}}_{d})^{T}$ is the random ancillary matrix with observed realization $\boldsymbol{A} = (\boldsymbol{a}_{1},\ldots,\boldsymbol{a}_{d})^{T}$. To derive \eqref{eq:RelInfo} the fact that ${\rm{E}}[\dot{l}_{\boldsymbol{\mathcal{Y}},\boldsymbol{\mathcal{T}}_{\negmedspace i}}|\boldsymbol{\mathcal{A}}_{i},\xi] = 0$ was used; see the proof of Proposition \ref{prop:CCRLB} for details [supplemental materials]. The dependence of  $\boldsymbol{H}_{\boldsymbol{A}}$ on $\boldsymbol{\theta}$ is omitted when the meaning is clear. See \cite{Fish:TwoN:1934}, \cite{Basu:Reco:1969}, \cite{Ghos:Reid:Fras:Anci:2010} and \cite{lane2021optimal} for additional details regarding the Fisher information in the sample versus the Fisher information in the relevant subset. Illustrative examples are provided later in this section to help differentiate the measures of information.

An argument in favor of $\boldsymbol{F}_{\xi}$ is that its inverse forms a lower bound on the variance of estimates of $\boldsymbol{\theta}$. A similar argument for $\boldsymbol{H}_{\boldsymbol{A}}$ is the following conditional CRLB.
\begin{proposition} \label{prop:CCRLB}
Under Condition \ref{cond:RegularExt} ${\rm{Var}}[\tilde{\boldsymbol{\theta}}|\boldsymbol{\mathcal{A}},\xi] \ge \boldsymbol{H}_{\boldsymbol{\mathcal{A}}}^{-1}$
for all $\xi\in\Xi_{n}$.
\end{proposition}
See the supplemental materials for proof. Noting that ${\rm{Var}}[\tilde{\boldsymbol{\theta}}|\boldsymbol{\mathcal{A}},\xi] \ge \boldsymbol{F}_{\xi}^{-1}$ is not guaranteed reinforces the value of $\boldsymbol{H}_{\boldsymbol{\mathcal{A}}}$ from a conditional perspective. Historically, the literature regarding conditioning on the relevant subset has primarily focused on post-data conditional inference [\cite{Fish:TwoN:1934,Cox:Some:1958,Basu:Reco:1969,McCu:Cond:1992,Efro:Hink:Asse:1978,Ghos:Reid:Fras:Anci:2010}]. Before an experiment is conducted the relevant subset is unknown and conditional properties are an insufficient design motivator. It is more important to assess a designs unconditional properties; e.g. does a design improve the unconditional variance of the estimates of $\boldsymbol{\theta}$. To this end Proposition \ref{prop:CCRLB} is used to derive the following tighter lower bound on the unconditional variance of an estimate of $\boldsymbol{\theta}$. 
\begin{proposition} \label{prop:RSLB}
Under Condition \ref{cond:RegularExt}
\begin{align} \label{eq:RSLB}
    {\rm{Var}}[\tilde{\boldsymbol{\theta}}|\xi] \ge {\rm{E}}[\boldsymbol{H}_{\boldsymbol{\mathcal{A}}}^{-1}|\xi] \ge \boldsymbol{F}_{\xi}^{-1}
\end{align}
for all $\xi\in\Xi_{n}$.
\end{proposition} 
See the supplemental materials for proof. The term ${\rm{E}}[\boldsymbol{H}_{\boldsymbol{\mathcal{A}}}^{-1}|
\xi]$ will be referred to as the relevant subset lower bound (RSLB). Proposition \ref{prop:RSLB} shows that the RSLB is tighter, in Loewner order, than the CRLB. A notable case where equality is obtained for both $\ge$ signs in \eqref{eq:RSLB} is the normal linear model with constant variance.


\begin{remark} 
The condition that $\boldsymbol{b}_{\boldsymbol{A}} (\boldsymbol{\theta}) = \boldsymbol{\theta}$ for all $\boldsymbol{A}$ is stronger than the condition that $\boldsymbol{v} (\boldsymbol{\theta}) = \boldsymbol{\theta}$ required for the CRLB. As was the case for the traditional CRLB it is reasonable to expect that most estimates of interest have negligible conditional bias. This discussion suggests that ${\rm{E}}[\boldsymbol{H}_{\boldsymbol{\mathcal{A}}}^{-1}|\xi]$ is generally the dominant term in the RSLB just as $\boldsymbol{F}_{\xi}^{-1}$ is the dominant term in the CRLB. 
\end{remark}

\begin{remark} \label{rem:OtherLB}
There exist tighter alternatives to the CRLB [\cite{Bhat:OnSo:1948,VanT:Dete:1968,Bobr:Maye:Zaka:Some:1987}, and others]. A search for the bound \eqref{eq:RSLB} in the existing literature did not yield any similar expressions. Due to the simplicity of its derivation it is possible, perhaps even likely, that its existence has been noted. One potential reason \eqref{eq:RSLB} may not be well known that a closed form expression for the distribution of $\boldsymbol{H}_{\boldsymbol{\mathcal{A}}}^{-1}$ may not exist which makes computing its expectation a challenge. This is examined in detail in Section \ref{subsec:GenSet}. Conversely, the CRLB is relatively straightforward to compute.
\end{remark}

In light of Remark \ref{rem:OtherLB} it is important to note that the RSLB is not presented as a post-data alternative to the CRLB. Instead, it is meant to demonstrate the connection between efficiency and the relevant subset and, more importantly for the current work, that the RSLB represents a theoretical benchmark for assessing the quality of candidate designs based on the conventional wisdom that improving a tight lower bound on the variance improves inference. The value of the RSLB as a benchmark is more precisely described in Section \ref{sec:RSD}.

The remainder of this section derives the relevant subset, the information measures and the lower bounds in two distinct settings. The first setting is an illustrative example where the MLE attains the RSLB but not the CRLB. This first example provides intuition for the RSLB but has limited practical value. The second setting includes a broad class of models and represents the general framework for the present work. 

\subsection{Illustrative Setting} \label{subsec:NIG}
Suppose the $j$th response is conditionally normal with random ancillary error variance $\mathcal{A}(j)$. Specifically, $\mathcal{Y}(j)|\mathcal{A}(j) = a(j) \sim N[\eta(j),1/a(j)]$, where $\eta(j) =  \boldsymbol{{\rm{f}}}^{T}[\boldsymbol{x}(j)]\boldsymbol{\theta}$ is restricted to be linear with respect to $\boldsymbol{\theta}$ for $j=1,\ldots,n$. The log-likelihood of $\eta_{i}$ is
\begin{align} \label{eq:etalik}
    l(\eta_{i}|\boldsymbol{y},\boldsymbol{t}_{i}) \propto \sum_{j=1}^{n} t_{i}(j) a(j) [y(j) - \eta_{i}]^{2}.
\end{align}
From its definition the Fisher information in relevant subset on $\eta_{i}$ is 
$h_{\boldsymbol{a}_{i}} = \sum_{j} t_{i}(j) a(j)$, where $\boldsymbol{a}_{i} = \{a(j):t_{i}(j) = 1, j= 1,\ldots,n\}$. The MLE, $\hat{\eta}_{i}$, is the value of $\eta_{i}$ that minimizes \eqref{eq:etalik} and can be obtained via a weighted least squares approach as $\hat{\eta}_{i} = h_{\boldsymbol{a}_{i}}^{-1} \sum_{j} t_{i}(j) a(j) y(j)$. The MLE is a linear combination of conditionally independent normal random variables which implies that $\hat{\eta}_{i}|\boldsymbol{a}_{i}$ is normally distributed with mean $\eta_{i}$ and variance $h_{\boldsymbol{a}_{i}}^{-1}$. 

To extend this to $\boldsymbol{\theta}$, note that $\log L(\boldsymbol{\theta}|\boldsymbol{y},\boldsymbol{T}) \propto \sum_{i} h_{\boldsymbol{a}_{i}} [\hat{\eta}_{i} - \boldsymbol{\theta}^{T}\boldsymbol{{\rm{f}}}(\boldsymbol{x}_{i})]^{2}$
and the MLE of $\boldsymbol{\theta}$ is also the solution to a weighted least squares problem; i.e. $\hat{\boldsymbol{\theta}} = \boldsymbol{H}_{\boldsymbol{A}}^{-1} {\rm{\textbf{F}}}^{T}\boldsymbol{K}_{\boldsymbol{A}}\hat{\boldsymbol{\eta}}$, where $\hat{\boldsymbol{\eta}} = (\hat{\eta}_{1},\ldots,\hat{\eta}_{d})^{T}$, $\boldsymbol{A} = (\boldsymbol{a}_{1},\ldots,\boldsymbol{a}_{d})^{T}$,  $\boldsymbol{K}_{\boldsymbol{A}}={\rm{diag}}(h_{\boldsymbol{a}_{1}},\ldots,h_{\boldsymbol{a}_{d}})$, ${\rm{\textbf{F}}}$ is a $d\times p$ matrix with $i$th row $\textbf{f}^{T}(\boldsymbol{x}_{i})$ and $\boldsymbol{H}_{\boldsymbol{A}} = {\rm{\textbf{F}}}^{T}\boldsymbol{K}_{\boldsymbol{A}}{\rm{\textbf{F}}}$. Since $\hat{\boldsymbol{\theta}}$ is a linear combination of conditionally normal random variables it follows that $\hat{\boldsymbol{\theta}}|\boldsymbol{A}$ is normally distributed with mean $\boldsymbol{\theta}$ and variance $\boldsymbol{H}_{\boldsymbol{A}}^{-1}$. This directly implies that $\mbox{Var}[\hat{\boldsymbol{\theta}}|\xi] = {\rm{E}}[\boldsymbol{H}_{\boldsymbol{\mathcal{A}}}^{-1}|\xi] \ge \boldsymbol{F}_{\xi}^{-1}$ which confirms that the MLE attains the RSLB but not the CRLB. 

\subsubsection{Example: Normal Inverse-Gamma}

Return to the two treatment example in Section \ref{sec:Intro} and let $\mathcal{A}(j) \sim {\rm{Gamma}}(\alpha,\beta)$, where $\alpha,\beta>0$. In the notation of model \eqref{eq:model} $\boldsymbol{{\rm{f}}}^{T}(\boldsymbol{x}_{i}) = \boldsymbol{x}_{i}$, $\eta_{i} = \boldsymbol{x}_{i}^{T}\boldsymbol{\theta}$ and the design space $\mathscr{X}$ contains two points $\boldsymbol{x}_{1} = (1,1)^{T}$ and $\boldsymbol{x}_{2} = (1,0)^{T}$ corresponding to treatments 1 and 2, respectively. Under the present scenario, $h_{\boldsymbol{a}_{i}}$ is the sum of gamma distributed random variables which implies that $h_{\boldsymbol{a}_{i}} \sim {\rm{Gamma}}(nw_{i}\alpha,\beta)$ and $h_{\boldsymbol{a}_{i}}^{-1} \sim \mbox{Inverse-Gamma}(nw_{i}\alpha,\beta)$. From the definition of the normal inverse-gamma (NIG) distribution it can be shown that $(\hat{\eta}_{i},h_{\boldsymbol{a}_{i}}^{-1}) \sim {\rm{NIG}}(\eta_{i},nw_{i}\alpha,\beta)$ [see \cite{Gelm:Carl:Ster:Baye:2014} ch. 2]. The MLE of $\Delta = \eta_{1} - \eta_{2}$, denoted $\hat{\Delta} = \hat{\eta}_{1} - \hat{\eta}_{2}$, has mean ${\rm{E}}[\hat{\Delta}|\xi] = \Delta$ and variance ${\rm{Var}}[\hat{\Delta}|\xi] = \beta[1/(\alpha nw_{1} - 1) +  1/(\alpha nw_{2} - 1)]$.   

Recall from Section \ref{sec:Intro} that $\rho$ represents the efficient allocation proportion. In the current setting the treatment variances are equal an $\rho = 1/2$; i.e. the design $\xi_{\rho}$ that assigns an equal number of observations to each treatment (i.e. $w_{1} = w_{2} = 1/2$) is efficient. In this illustration $\rho$ is known and there is no need to estimate it using the adaptive procedure described in the introduction. For the balanced design ${\rm{Var}}[\hat{\Delta}|\xi_{\rho}] = 2\beta/(n\alpha/2 - 1)$. Experiments require  $n>2/\alpha$ observations in order to ensure a finite variance. As shown above, the variance of $\hat{\Delta}$ and the RSLB coincide, i.e. ${\rm{Var}}[\hat{\Delta}|\xi_{\rho}] = {\rm{E}}[\boldsymbol{H}_{\boldsymbol{\mathcal{A}}}^{-1}|\xi_{\rho}]$. One interesting feature is that $\boldsymbol{F}_{\xi_{\rho}}^{-1}=\{{\rm{E}}[\boldsymbol{H}_{\boldsymbol{\mathcal{A}}}|\xi_{\rho}]\}^{-1} = \beta/(n\alpha)$ is always finite, regardless of sample size. 

\subsection{General Setting} \label{subsec:GenSet}

One common condition placed on additive error regression models, stated in \eqref{eq:model}, is that the errors are from the location family. The location family is characterized by the relation $f(y|\eta) = f(\varepsilon|0)$ and it is well established that there exists and exact ancillary, $\boldsymbol{A} = (\boldsymbol{a}_{1},\ldots,\boldsymbol{a}_{d})^{T}$, where $\boldsymbol{a}_{i} = [y_{i(2)} - y_{i(1)},\ldots,y_{i(n_{i})} - y_{i(n_{i}-1)}]^{T}$ and $y_{i(j)}$ is the $j$th order statistic of $\boldsymbol{y}_{i}$. This section outlines the advantages of the location condition for the current work.


At the beginning of this section the Fisher information in the relevant subset on $\boldsymbol{\theta}$, $\boldsymbol{H}_{\boldsymbol{A}}$, was constructed from the Fisher information in the relevant subset on $\eta_{i}$, $h_{\boldsymbol{a}_{i}}$. Instead of having to work with the $p$-dimensional parameter, $\boldsymbol{\theta}$, each of the one-dimensional parameters, $\eta_{1},\ldots,\eta_{d}$, can be examined separately. Let $\overline{l}(\eta_{i},\hat{\eta}_{i}) = l(\eta_{i}|\boldsymbol{y},\boldsymbol{t}_{i}) - l(\hat{\eta}_{i}|\boldsymbol{y},\boldsymbol{t}_{i})$; \cite{Fish:TwoN:1934} famously demonstrated that
\begin{align} \label{eq:GenCond}
    p(\hat{\eta}_{i}|\boldsymbol{a}_{i},\eta_{i}) = c_{\boldsymbol{a}_{i}}e^{\overline{l}(\eta_{i},\hat{\eta}_{i})},
\end{align}
where $c_{\boldsymbol{a}_{i}}$ is a normalizing constant that ensures the right hand side integrates to one. It is important to point out that it is not necessary to compute the ancillary statistic for the methods proposed in this work. What is required is that $h_{\boldsymbol{a}_{i}} = {\rm{E}}[i_{\boldsymbol{y},\boldsymbol{t}_{i}}|\boldsymbol{a}_{i},\xi]$ can be computed using analytical or numerical methods; examples are provided below to illustrate a useful approach to obtain this quantity. Once $h_{\boldsymbol{a}_{i}}$ has been computed $\boldsymbol{H}_{\boldsymbol{A}}$ can be computed using \eqref{eq:RelInfo}.

A second advantage of the location setting is that it allows for an invariant form of information to be considered. Denote the invariant observed information as $i_{\boldsymbol{y},\boldsymbol{t}_{i}}^{(\phi)} = i_{\boldsymbol{y},\boldsymbol{t}_{i}}/\mu$, where the superscript $(\phi)$ is used to recognized that it is invariant to any one-to-one transformation, $\phi(\eta_{i})$. Noting that \eqref{eq:GenCond} is invariant with respect to one-to-one transformations [\cite{Barn:Cox:Infe:1994} ch 6] yields
\begin{align}
    g_{\boldsymbol{a}_{i}} =  {\rm{E}}[i_{\boldsymbol{y},\boldsymbol{t}_{i}}^{(\phi)}|\boldsymbol{a}_{i},\xi] = \mu h_{\boldsymbol{a}_{i}},
\end{align}
where $g_{\boldsymbol{a}_{i}}$ is the invariant information in the relevant subset. Two practical examples are provided below. The design proposed in the next section depends only on $g_{\boldsymbol{a}_{1}},\ldots,g_{\boldsymbol{a}_{d}}$ and is thus invariant to one-to-one transformations of $\eta_{i}$.  

\begin{remark}
The challenge with computing ${\rm{E}}[\boldsymbol{H}_{\boldsymbol{\mathcal{A}}}^{-1}|\xi]$, alluded to in Remark \ref{rem:OtherLB} can now be made more explicit. A closed form expression for the distribution of the inverse of $\boldsymbol{H}_{\boldsymbol{A}}$ likely does not exist, which makes obtaining its expectation via analytical means impossible. Monte Carlo methods can be used in place of exact solutions, but such methods are computationally expensive. As previously stated, this discussion indicates that the RSLB may not be a suitable replacement to the CRLB as a post-data variance lower bound. One critical feature of the adaptive design proposed in Section \ref{sec:RSD} is that it does not require ${\rm{E}}[\boldsymbol{H}_{\boldsymbol{\mathcal{A}}}^{-1}|\xi]$ to be computed.
\end{remark}

\subsubsection{Example: Generalized Normal} \label{subsubsec:GND}

The generalized normal distribution (GND) has PDF $f(y|\eta) \propto e^{-\zeta^{-1}|(y - \eta)/\tau|^{\zeta}}$ with $y,\eta\in\mathbb{R}$ and $\zeta,\tau>0$. Special cases include the double exponential distribution for $\zeta = 1$, the normal distribution for $\zeta = 2$ and the uniform distribution as $\zeta\rightarrow\infty$. The log-likelihood of $\eta_{i}$ is $l(\eta_{i}|\boldsymbol{y},\boldsymbol{t}_{i}) = -\zeta^{-1} \sum_{j = 1}^{n}t_{i}(j) |[y(j) - \eta_{i}]/(\zeta\tau)|^{\zeta}$ and the observed information from subject $j$ is $i_{y(j)} = (\zeta - 1)\{[y(j) - \eta(j)]/\tau\}^{\zeta - 2}/\tau^{2}$ which has expectation $\mu =\zeta^{2(1-1/\zeta)} \Gamma(2 - 1/\zeta) / [2\tau^{2}\Gamma[1/\zeta]]$, where $\eta(j) = \eta[\boldsymbol{x}(j),\boldsymbol{\theta}]$. 

To compute $h_{\boldsymbol{a}_{i}}$ use the change of variable, $v = \hat{\eta}_{i} - \eta_{i}$ to obtain
\begin{align}
    h_{\boldsymbol{a}_{i}} &= \int i_{\boldsymbol{y},\boldsymbol{t}_{i}}(\hat{\eta}_{i} - v) e^{\overline{l}(\hat{\eta}_{i} - v,\hat{\eta}_{i})} dv / \int e^{\overline{l}(\hat{\eta}_{i} - v,\hat{\eta}_{i})} dv 
\end{align}
which can be computed using numeric integration. The dependence of $i_{\boldsymbol{y},\boldsymbol{t}_{i}}(\eta_{i})$ on $\eta_{i}$ has been made explicit for clarity. The invariant measure is recovered by dividing by $\mu$; i.e. $g_{\boldsymbol{a}_{i}} = h_{\boldsymbol{a}_{i}}/\mu$. This assumes that $\mu$ is known. In the event that $\mu$ is unknown it can be replaced with its MLE $\hat{\mu}$. 

\subsubsection{Example: Cauchy} \label{subsubsec:Cauchy}

The Cauchy distribution has PDF $f(y|\eta) \propto 1/\{1 + [(y - \eta)/\tau]^{2}\}$ for $y,\eta\in\mathbb{R}$ and $\tau>0$ which implies that $l(\eta_{i}|\boldsymbol{y},\boldsymbol{t}_{i}) = -\sum_{j = 1}^{n}t_{i}(j) \log\{1 + [u^{2}(j)/\tau^{2}]\}$, $i_{y(j)} = 2[\tau^{2} - u^{2}(j)]/[\tau + u^{2}(j)]^{2}$ and $\mu = 1/(2\tau^{2})$, where $u(j) = y(j) - \eta(j)$. The steps from the preceding example can be repeated to obtain $g_{\boldsymbol{a}_{i}}$.

\section{Relevant Subset Design} \label{sec:RSD}
As stated previously, the RSLB is difficult to compute and may not be widely accepted as a post-data variance bound replacement for the CRLB. In this section it is shown that the RSLB has significant value from a design perspective. The following additional condition is required for the adaptive design proposed in this work.

\begin{condition}\label{cond:C2}
(1) The error distribution is a member of the location family; (2) $\gamma^{2} = {\rm{Var}}[i_{\mathcal{Y}}^{(\phi)}] < \infty$, where $\gamma = [(\nu_{02}\nu_{20} - \nu_{11})/\nu_{20}^{3}]^{1/2}$ and $\nu_{kl} = {\rm{E}}[\dot{l}_{\mathcal{Y}}^{k}(\ddot{l}_{\mathcal{Y}} + {\rm{E}}[\dot{l}_{\mathcal{Y}}^{2}])^{l} ]$ (2) E$[(\hat{\eta}_{i} - \eta_{i})^{2}]<\infty$ (3) $l_{y}^{(\cdot k)} - (\partial^{k}/\partial\eta^{k})\log f(y|\eta)$ is bounded for $k=1,\ldots,4$; and (4) $(\partial^{3}/\partial w_{i}w_{j}w_{k})\boldsymbol{M}_{\xi}^{-1}$ is bounded for all $i,j,k = 1,\ldots,d$, where $\boldsymbol{M}_{\xi} = \boldsymbol{F}_{\xi}/(n\mu)$.
\end{condition}

Historically, designs that minimize the CRLB, i.e. ``minimize'' $\boldsymbol{F}_{\xi}^{-1}$, have been considered \emph{efficient}; note, $\boldsymbol{F}_{\xi}^{-1}$ is a matrix and cannot be minimized in general. As described in Section \ref{sec:LB} the CRLB is not necessarily tight. In order to describe the efficiency guaranteed by designs that are efficient with respect to the CRLB consider the following Proposition.
\begin{proposition} \label{prop:LBexp}
Under Condition \ref{cond:RegularExt} and \ref{cond:C2} 
\begin{align} \label{eq:Vdiff}
    \lim_{n\rightarrow\infty} n^{2}\boldsymbol{c}^{T}({\rm{E}}[\boldsymbol{H}_{\boldsymbol{\mathcal{A}}}^{-1}|\xi] - \boldsymbol{F}_{\xi}^{-1})\boldsymbol{c} &= \mu\lim_{n\rightarrow\infty} n^{-1} tr\{\boldsymbol{D}\left( {\rm{Var}}[\boldsymbol{u}_{\boldsymbol{\mathcal{A}}}|\xi] + {\rm{Var}}[\boldsymbol{v}_{\boldsymbol{\mathcal{A}}}|\xi]] \right) \}
\end{align}
for all $\xi\in\Xi_{n}$, where $\boldsymbol{u}_{\boldsymbol{A}} = (u_{\boldsymbol{a}_{1}},\ldots,u_{\boldsymbol{a}_{d}})^{T}$, $u_{\boldsymbol{a}_{i}} =  g_{\boldsymbol{a}_{i}} - w_{i}\sum_{i=1}^{d}g_{\boldsymbol{a}_{i}}$, $\boldsymbol{v}_{\boldsymbol{A}} = (v_{\boldsymbol{a}_{1}},\ldots,v_{\boldsymbol{a}_{d}})^{T}$, $v_{\boldsymbol{a}_{i}} = w_{i}(\sum_{i=1}^{d}g_{\boldsymbol{a}_{i}} - n)$, $\boldsymbol{D}$ is a $d\times d$ symmetric matrix with $ij$th element equal to $r_{i}(\boldsymbol{c})r_{j}(\boldsymbol{c})s_{ij}$, $r_{i}(\boldsymbol{c}) = \boldsymbol{\dot{\eta}}^{T}(\boldsymbol{x}_{i}) \boldsymbol{M}_{\xi}^{-1}\boldsymbol{c}$, $s_{ij} = \boldsymbol{\dot{\eta}}^{T}(\boldsymbol{x}_{i})\boldsymbol{M}_{\xi}^{-1}\boldsymbol{\dot{\eta}}(\boldsymbol{x}_{j})$ and $\boldsymbol{\dot{\eta}}(\boldsymbol{x}_{i}) = (\partial/\partial\boldsymbol{\theta})\eta(\boldsymbol{x}_{i},\boldsymbol{\theta})$.
\end{proposition}  

In the proof  of Proposition \ref{prop:LBexp}, provided in the supplemental materials, it is shown that $n^{-1}{\rm{Var}}[\boldsymbol{v}_{\boldsymbol{\mathcal{A}}}|\xi] \rightarrow \gamma^{2}\boldsymbol{w}\boldsymbol{w}^{T}$ and $n^{-1}{\rm{Var}}[\boldsymbol{u}_{\boldsymbol{\mathcal{A}}}|\xi] \rightarrow \gamma^{2}\boldsymbol{W}(\boldsymbol{I}_{d} - \boldsymbol{w}\boldsymbol{w}^{T})$ as $n\rightarrow\infty$, where $\boldsymbol{W} = {\rm{diag}}(\boldsymbol{w})$. This implies that the right hand side of \eqref{eq:Vdiff} converges to a constant as $n\rightarrow\infty$. Propositions \ref{prop:RSLB} and \ref{prop:LBexp} indicate that a design that is efficient with respect to  $\boldsymbol{F}_{\xi}^{-1}$ is efficient to the order $O(n^{-1})$. Formally, a design $\xi_{1}\in\Xi_{n}$ \textbf{is} \emph{first order efficient} if there \textbf{does not exist} a design $\xi_{2}\in\Xi_{n}$ such that
\begin{align} \label{eq:FiniteEff}
    n\boldsymbol{c}^{T}[\boldsymbol{F}_{\xi_{1}}^{-1} - \boldsymbol{F}_{\xi_{2}}^{-1}]\boldsymbol{c} > 0
\end{align}
for all nonzero $\boldsymbol{c}$. Note $\boldsymbol{c}^{T}(\boldsymbol{A} - \boldsymbol{B})\boldsymbol{c} > 0$ for all nonzero $\boldsymbol{c}$ implies $\boldsymbol{A}>\boldsymbol{B}$ in Loewner order. Not all designs satisfying this definition of efficiency have practical value; however, if efficiency is an objective then the design used in the experiment should.

Intuitively, a designs efficiency is improved if it minimises the RSLB, i.e. ``minimizes'' ${\rm{E}}[\boldsymbol{H}_{\boldsymbol{\mathcal{A}}}^{-1}|\xi]$. This in combination with Proposition \ref{prop:LBexp} motivates a higher order definition of efficiency; specifically, a design $\xi_{1}\in\Xi_{n}$ \textbf{is not} \emph{second order efficient} if there \textbf{exists} a design $\xi_{2}\in\Xi_{n}$ such that
\begin{align} \label{eq:SecondEff}
    \lim_{n\rightarrow\infty} n^{2}\boldsymbol{c}^{T}( {\rm{E}}[H_{\boldsymbol{\mathcal{A}}}^{-1}|\xi_{2}] -  {\rm{E}}[H_{\boldsymbol{\mathcal{A}}}^{-1}|\xi_{1}])\boldsymbol{c} > 0
\end{align}
for all nonzero $\boldsymbol{c}$. This is a reversal of the previous definition. Equation \eqref{eq:FiniteEff} describes a method to determine if a design \textbf{is} first order efficient; conversely, \eqref{eq:SecondEff} describes a method to determine if a design \textbf{is not} second order efficient. Despite its limitation \eqref{eq:SecondEff} is a useful tool to eliminate inefficient designs. This definition is used to confirm the main result of this work, that \textbf{no} \emph{a priori} design is second order efficient. 

Proposition \ref{prop:LBexp} also provides a motivation for an adaptive design. The difference between the RSLB and CRLB, is dominated by the variances of $\boldsymbol{u}_{\boldsymbol{A}}$ and $\boldsymbol{v}_{\boldsymbol{A}}$. In the next section an adaptive approach is derived that eliminates the variability associated with $\boldsymbol{u}_{\boldsymbol{A}}$; thus reducing the gap between the RSLB and CRLB and increasing efficiency. Section \ref{sec:Main} gives a rigorous treatment of this result. It is important to note that $\boldsymbol{u}_{\boldsymbol{A}}$ and $\boldsymbol{v}_{\boldsymbol{A}}$ are functions of the invariant information in the relevant subset $g_{\boldsymbol{a}_{1}},\ldots,g_{\boldsymbol{a}_{d}}$.

\subsection{Randomized Relevant Subset Design}

A sequential experiment consists of a series of runs, $r = 1,\ldots,\mathcal{R}_{\max}$, where each run has a potentially random number, $\mathcal{N}\{r\}$, of independent observations. While the total number of runs is random the total sample size, $n$, is fixed. The notation $\{r\}$ indicates a variable corresponds to run $r$; whereas $\{1;r\}$ corresponds to a variable that incorporates the data from all runs up to and including run $r$. The curly bracket is used to distinguish the $r$th run from the $j$th observation, which is denoted with a parentheses. For example, $y(j)$ is the response of observation $j$; $\boldsymbol{y}\{r\} = [y(\mathcal{N}\{1;r-1\}+1),\ldots,y(\mathcal{N}\{1;r\})]^{T}$ are the responses of the observations in run $r$; and $\boldsymbol{y}\{1;r\} = (\boldsymbol{y}^{T}\{1\},\ldots,\boldsymbol{y}^{T}\{r\})^{T}$ are the responses from all observations from the first $r$ runs, where $\mathcal{N}\{1;r\} = \sum_{s=1}^{r} \mathcal{N}\{s\}$ is the number of observations in the first $r$ runs. As a second example, $i_{y(j)}^{(\phi)}$ is the invariant observed information for the $j$th observation;
\begin{align} \label{eq:air}
i_{\boldsymbol{y}\{r\},\boldsymbol{t}_{i}\{r\}}^{(\phi)} = \sum_{j=\mathcal{N}\{1;r-1\}+1}^{\mathcal{N}\{1;r\}} t_{i}(j)i_{y(j)}^{(\phi)}
\end{align}
is the invariant observed information from run $r$; and $i_{\boldsymbol{y}\{1;r\},\boldsymbol{t}_{i}\{1;r\}}^{(\phi)} = \sum_{s=1}^{r} i_{\boldsymbol{y}\{r\},\boldsymbol{t}_{i}\{r\}}^{(\phi)}$ is the total invariant observed information from all runs up to and including run $r$. 

Adaptive designs are characterized by the use of the data from the preceding $r$ runs to determine the design of run $r+1$. In this work a method is proposed that uses the relevant subset from the first $r$ runs to determine the probability of assigning each observation in run $r+1$ to the $i$th support point, denoted $\mathcal{P}_{i}\{r+1\} = P\{\mathcal{T}_{i}(j)=1| \boldsymbol{\mathcal{A}}\{1;r\}\}$, and the sample size of run $r+1$, $\mathcal{N}\{r+1\}$. As described $\mathcal{P}_{i}\{r+1\}$ and $\mathcal{N}\{r+1\}$ are $\mathcal{A}\negmedspace\{1;r\}$ measurable.  
To develop the proposed design, consider the first two runs of experiment where $\mathcal{N}\{1\}=n\{1\}$ and $\mathcal{N}_{i}\{1\} = w_{i}n\{1\}$ are pre-specified constants and it is desired to determine the sample size of the second run, $\mathcal{N}\{2\}$, and the probability of assigning each observation in the second run to the $i$th support point, $\mathcal{P}_{i}\{2\}$. 

The most important quantity in the proposed design is $u_{\boldsymbol{\mathcal{A}}_{i}}$ defined as $u_{\boldsymbol{a}_{i}} =  g_{\boldsymbol{a}_{i}} - w_{i}\sum_{i=1}^{d}g_{\boldsymbol{a}_{i}}$ in Proposition \ref{prop:LBexp}, see Section \ref{subsec:GenSet} for details on its computation. For the derivation of the adaptive design proposed in this section the following alternative expression is informative
\begin{align}
u_{\boldsymbol{\mathcal{A}}_{i}\{1;r\}} 
&= {\rm{E}}\left[\left. \mathcal{S}_{i}\{1;r\} \right| \boldsymbol{\mathcal{A}}_{i}\{1;r\},\xi \right],
\end{align}
where $\mathcal{S}_{i}\{1;r\} = \sum_{s=1}^{r}\mathcal{S}_{i}\{s\}$, $\mathcal{S}_{i}\{r\} = \sum_{j=\mathcal{N}\{1;r-1\}+1}^{\mathcal{N}\{1;r\}}s_{\mathcal{Y}(j)}$ and $s_{\mathcal{Y}(j)} = \{\mathcal{T}_{i}(j) -  w_{i}\} i_{\mathcal{Y}(j)}^{(\phi)}$. 


Consider the expectation of $u_{\boldsymbol{\mathcal{A}}_{i}\{1;2\}}$ conditional on $\boldsymbol{\mathcal{A}}\{1\}$
\begin{align} 
 {\rm{E}}\left[u_{\boldsymbol{\mathcal{A}}_{i}\{1;2\}}| \boldsymbol{\mathcal{A}}\{1\}, \xi \right] &= {\rm{E}}\left[\left.\mathcal{S}_{i}\{1\} + \mathcal{S}_{i}\{2\} \right| \boldsymbol{\mathcal{A}}\{1\} , \xi \right] \\
 &=  u_{\boldsymbol{\mathcal{A}}_{i}\{1\}} + \sum_{j=n\{1\}+1}^{\mathcal{N}\{1;2\}} {\rm{E}}\left[\left. s_{\mathcal{Y}(j)}  \right| \boldsymbol{\mathcal{A}}\{1\}, \xi \right] \\
 &= u_{\boldsymbol{\mathcal{A}}_{i}\{1\}} + \mathcal{N}\{2\} (\mathcal{P}_{i}\{2\} - w_{i})  \label{eq:uai2}
\end{align}
The first line is from the definition of $u_{\boldsymbol{\mathcal{A}}_{i}\{1;2\}}$ and the fact that $\boldsymbol{\mathcal{A}}_{i}\{1\}\subseteq\boldsymbol{\mathcal{A}}_{i}\{1;2\}$ [see the proof of Lemma \ref{lem:TwoStage} for proof]; the second is obtained from the definitions of $u_{\boldsymbol{\mathcal{A}}_{i}\{1\}}$ and $\mathcal{S}_{i}\{2\}$ and by recognizing that $\mathcal{N}\{1;2\}$ is $\boldsymbol{\mathcal{A}}_{i}\{1\}$ measurable; the third line follows from
\begin{align}
    {\rm{E}}[s_{\mathcal{Y}(j)}| \boldsymbol{\mathcal{A}}\{1\},\xi] &=  {\rm{E}}[i_{\mathcal{Y}(j)}^{(\phi)}|\mathcal{T}_{i}(j)](P\{\mathcal{T}_{i}(j)=1| \boldsymbol{\mathcal{A}}\{1\}\}  -  w_{i} ) = \mathcal{P}_{i}\{2\} - w_{i}
\end{align} 
for $j = n\{1\}+1,\ldots,\mathcal{N}\{2\}$.

Consider a careful selection of $\mathcal{N}\{2\}$ and $\mathcal{P}_{i}\{2\}$. Set the right hand side of \eqref{eq:uai2} equal to 0 and solve for $\mathcal{P}_{i}\{2\}$ to obtain
\begin{align} \label{eq:pid2}
     \mathcal{P}_{i}\{2\} = w_{i} - u_{\boldsymbol{\mathcal{A}}_{i}\{1\}}/\mathcal{N}\{2\}.
\end{align}
As defined, $\mathcal{P}_{i}\{2\}$ is a probability and it is implicitly required that $\mathcal{P}_{i}\{2\} \ge 0$. This is only guaranteed if $\mathcal{N}\{2\}$ is sufficiently ``large". This is ensured by setting the right hand side of \eqref{eq:pid2} to zero, solving to obtain $\mathcal{N}_{i}\{2\} = u_{\boldsymbol{\mathcal{A}}_{i}\{1\}}/w_{i}$ and defining
\begin{align} \label{eq:u2}
  \mathcal{N}\{2\} = \lceil\max\{\mathcal{N}_{1}\{2\},\ldots,\mathcal{N}_{d}\{2\}\}\rceil, 
\end{align} 
where $\lceil \cdot \rceil$ is the ceiling operator. The ceiling operation ensures $\mathcal{N}\{2\}$ is an integer and the $\max$ ensures that all $\mathcal{P}_{i}\{2\} \ge 0$. Plugging this selection of $\mathcal{N}\{2\}$ and $\mathcal{P}_{i}\{2\}$ into \eqref{eq:uai2} yields ${\rm{E}}[u_{\boldsymbol{\mathcal{A}}_{i}\{1;2\}}| \boldsymbol{\mathcal{A}}\{1\}]=0$. To appreciate the significance of this accomplishment consider an application of law of total variance which states that
\begin{align}
    {\rm{Var}}[u_{\boldsymbol{\mathcal{A}}\{1;2\}}|\xi] &= {\rm{Var}}[{\rm{E}}[u_{\boldsymbol{\mathcal{A}}\{1;2\}}| \boldsymbol{\mathcal{A}}\{1\},\xi]\mid\xi] + {\rm{E}}[{\rm{Var}}[u_{\boldsymbol{\mathcal{A}}\{1;2\}}| \boldsymbol{\mathcal{A}}\{1\},\xi]\mid\xi] \\
    &= {\rm{E}}[{\rm{Var}}[u_{\boldsymbol{\mathcal{A}}\{1;2\}}| \boldsymbol{\mathcal{A}}\{1\},\xi]\mid\xi]. \label{eq:VarUA12}
\end{align}
The above states that using the proposed adaptive procedure forces the expectation of $u_{\boldsymbol{\mathcal{A}}_{i}\{1;2\}}$, conditional on $\boldsymbol{\mathcal{A}}\{1\}$, to equal zero which in turn reduces its variance. In fact as shown in the below lemma variance of $u_{\boldsymbol{\mathcal{A}}_{i}\{1;2\}}$ is eliminated for large $n\{1\}$.
\begin{lemma} \label{lem:TwoStage}
Under Condition \ref{cond:RegularExt} and $\mathcal{N}\{2\}$ and $\mathcal{P}_{i}\{2\}$ as defined above, if $n\{1\}\rightarrow\infty$ then
\begin{align}
    \lim_{n\{1\}\rightarrow\infty} (n\{1\})^{-1} {\rm{Var}}[u_{\boldsymbol{\mathcal{A}}\{1;2\}}|\xi] = 0.
\end{align}
\end{lemma}
See section \ref{sec:Tech} for proof. Lemma \ref{lem:TwoStage} represents a significant improvement compared to a deterministic design where it was found that $(n\{1;2\})^{-1}{\rm{Var}}[\boldsymbol{u}_{\boldsymbol{\mathcal{A}}\{1;2\}}|\xi] =\gamma^{2}\boldsymbol{W}(\boldsymbol{I}_{d} - \boldsymbol{w}\boldsymbol{w}^{T})$. Recalling Proposition \ref{prop:LBexp} this implies that the adaptive procedure considerably reduces the RSLB as compared to a \emph{a priori} deterministic design.  

The two-run demonstration describes the first step of a more general approach where this procedure is repeated until the total sample size, $n$, is exhausted. The general approach, described below, is referred to as the Randomized Relevant Subset Design (RRSD).

\begin{algorithm} \label{alg:RRSD} Randomized Relevant Subset Design (RRSD)
\begin{enumerate}
    \item{Determine an \emph{a priori} design, $\xi\in\Xi_{n}$ with support $\boldsymbol{X} = (\boldsymbol{x}_{1},\ldots,\boldsymbol{x}_{d})^{T}$ and allocation weights $\boldsymbol{w} = (w_{1},\ldots,w_{d})^{T}$.}
    \item{Allocate the first $n\{1\}$, observations according to design $\xi$. If $n\{1\}w_{i}$ are integers for allocate exactly $n\{1\}w_{i}$ to treatment $i$; otherwise allocate with probability $\mathcal{P}_{i}\{1\} = w_{i}$ to treatment $i$.}
    \item{Compute $\mathcal{N}_{{\rm{temp}}}\{r+1\} = \lceil \max\{\mathcal{N}_{1}\{r+1\},\ldots,\mathcal{N}_{d}\{r+1\}\}\rceil$, where $\mathcal{N}_{i}\{r+1\} = u_{\boldsymbol{\mathcal{A}}_{i}\{1;r\}}/w_{i}$.}
    \item{\textbf{If} $\mathcal{N}_{{\rm{temp}}}\{r+1\} \le n - \mathcal{N}\{1;r-1\}$ let $\mathcal{N}\{r+1\} = \mathcal{N}_{{\rm{temp}}}\{r+1\}$ and
    \begin{align} \label{eq:Pir}
        \mathcal{P}_{i}\{r+1\} = w_{i} - u_{\boldsymbol{\mathcal{A}}_{i}\{1;r\}}/\mathcal{N}\{r+1\};
    \end{align} 
    \textbf{else}, let $\mathcal{P}_{i}\{r+1\} = w_{i}$ and $\mathcal{N}\{r+1\} = n - \mathcal{N}\{1;r-1\}$. }
    \item{Allocate observations $j=\mathcal{N}\{1;r\}+1,\ldots,\mathcal{N}\{1;r+1\}$ to the $i$th support point with probability $\mathcal{P}_{i}\{r+1\}$.}
    \item{Repeat steps 3-6 until $\mathcal{N}\{1;r+1\} = n$.}
\end{enumerate}
\end{algorithm}
The \textbf{if-else} condition in step 4 is required to ensure the adaptive design does not result in an experiment with a sample size greater than $n$. Let $\check{\xi} = [\check{x}(1),\ldots,\check{x}(n)]^{T}$ represent the observed design following the RRSD with initial design $\xi$, where $\check{x}(j)$ is the support point of observation $j$ assigned by algorithm. Additionally, the dependence on runs $\{1;\mathcal{R}_{\max}\}$ is often omitted for variables corresponding to all $n$ observations in the experiment, e.g. $\boldsymbol{\mathcal{A}} = \boldsymbol{\mathcal{A}}\{1;\mathcal{R}_{\max}\}$. An ancillary random variable is marked with a check, $\boldsymbol{\mathcal{\check{A}}}_{\xi}$, indicates that it corresponds to a RRSD, where the subscript $\xi$ is used to denote that fact that the corresponding RRSD was initialized by design $\xi$. Ancillary random variables without a check, $\boldsymbol{\mathcal{A}}_{\xi}$, corresponds to an experiment with deterministic design $\xi$.  Estimates of $\boldsymbol{\theta}$ are functions of the observed design, a superscript will be used to indicate this dependence; e.g. $\tilde{\boldsymbol{\theta}}_{\xi}$ and $\tilde{\boldsymbol{\theta}}_{\check{\xi}}$ represent estimates following a deterministic design and the RRSD, respectively.

A note on the deterministic design, $\xi$, that \emph{initializes} Algorithm \ref{alg:RRSD}. The selection of this design can be based on any existing deterministic design strategy, factorial, optimal, minimax, etc. The only requirement is that it satisfies Definition \ref{def:FixedDesign}. The selection of $\xi$ is critical to the overall performance of the RRSD. It is shown in the Section \ref{sec:Main} that implementing the RRSD always increases the efficiency compared to the design used to initialize it. If $\xi$ is inefficient then the RRSD is likely to be inefficient compared to an alternative RRSD initialized with a more efficient deterministic design. It is critical to recognize that the proposed approach does not negate the importance of \emph{a priori} deterministic design strategies.

\subsection{Random A Priori Designs} \label{subsec:Random}
Up to this point only deterministic \emph{a priori} designs have been considered. As stated in Section \ref{sec:Intro}, the algorithm and main results of this work extend to random \emph{a priori} designs. The conceptual change from a deterministic to a random strategy is that in the latter the design $\xi$ is an observed realization from a probability measure $\pi$, which places positive probability on one or more elements of $\Xi_{n}$. The only alteration to the RRSD algorithm required is that the experimenter must first randomized by generating an observed design $\xi$ from $\pi$ and then adapt by initializing Algorithm \ref{alg:RRSD} with this design. The importance of this \emph{randomize then adapt} strategy is outlined in Section \ref{subsec:MainRandom}.

\subsubsection{Minimax Random Design} \label{subsubsubsec:Minmax}
As an illustration consider a slightly generalized version of the \cite{Wait:Wood:Mini:2021} Example 4.2.1 where $x\in\mathscr{X} = [-1,1]$, $\eta(x,\boldsymbol{\theta}) = \boldsymbol{{\rm{f}}}^{T}(x)\boldsymbol{\theta}$ and $\boldsymbol{{\rm{f}}}(x)=(1,x,x^{2})^{T}$. The stated objective is to minimize the maximum prediction variance over $\mathscr{X}$. The minimax deterministic design, denoted $\xi_{G}$, minimizes the $\max_{x\in\mathscr{X}} {\rm{Var}}[\boldsymbol{{\rm{f}}}^{T}(x)\hat{\boldsymbol{\theta}}]$ and is efficient with respect to this objective. If $n$ is divisible by 3 then $\xi_{G}$ places $n/3$ observations on each point -1, 0 and 1. If $n \;\mathrm{mod}\; 3 = 1$ then $\xi_{G}$ places $(n-1)/3$ observations on each support point and arbitrarily assigns the remaining observation to one of -1, 0 or 1. \cite{Wait:Wood:Mini:2021} show that a strategy where $(n-1)/3$ observations are placed on each support point and the remaining observation is randomly assigned, with probability 1/3, to -1, 0 or 1 outperforms the deterministic approach. Let $\pi_{G}$ represent the minimax randomized design.   

The RRSD is only guaranteed to improve the efficiency of the design used to initialize it; a RRSD initialized with $\xi_{G}$ is not necessarily more efficient than the random design $\pi_{G}$ and visa versa. If the objective is to improve $\pi_{G}$ one should first randomize by generating an observed design, $\xi$, from $\pi_{G}$ and then adapt by initializing the RRSD with this observed realization. This randomize then adapt strategy is guaranteed to be more efficient than $\pi_{G}$ [Section \ref{subsec:MainRandom}]. This example is considered further in the simulation study of Section \ref{sec:Examples}.

Under the previously established notation, $\check{\pi}_{G}$, $\boldsymbol{\mathcal{\check{A}}}_{\pi_{G}}$ and $\tilde{\boldsymbol{\theta}}_{\check{\pi}_{G}}$ are the observed design, an ancillary variable and an estimate of $\boldsymbol{\theta}$ following the RRSD with initial design $\pi_{G}$.

\section{Main Result} \label{sec:Main}


The adaptive algorithm was motivated by the RSLB, shown in Section \ref{sec:LB} to hold for any design $\xi\in\Xi_{n}$. This is appropriate to motivate the proposed design; however, it is insufficient to reach the desired conclusion, that adaptation improves efficiency. For this it is required to extend Proposition \ref{prop:RSLB} to the RRSD.
\begin{lemma} \label{lem:AdaptRSLB}
Under the conditions of Proposition \ref{prop:RSLB} and Condition \ref{cond:C2}
\begin{align}{\rm{Var}}[\tilde{\boldsymbol{\theta}}_{\check{\xi}}|\xi] \ge {\rm{E}}[H_{\boldsymbol{\mathcal{\check{A}}}_{\xi}}^{-1}|\xi] \ge \boldsymbol{F}_{\xi}^{-1}
\end{align}
for all $\xi\in\Xi_{n}$.
\end{lemma}
See section \ref{sec:Tech} for proof.  Note that the expectation and variance in the lemma are conditional on the initial design $\xi$ and not the observed design $\check{\xi}$. 


The second inequality in Lemma \ref{lem:AdaptRSLB} shows that the RRSD has the same CRLB as the deterministic design used to initialize it. Now that it has been shown that the same lower bounds hold for the RRSD its efficiency, relative to $\xi$, can be assessed. The main result of this work is obtained in the following theorem. 
\begin{theorem} \label{thm:MainRes}
Under Conditions \ref{cond:RegularExt} and \ref{cond:C2} 
    \begin{align} \label{eq:AdaptCoxLim}
    \lim_{n\rightarrow\infty} n^{2}\boldsymbol{c}^{T}({\rm{E}}[H_{\boldsymbol{\mathcal{A}}_{\xi}}^{-1}|\xi] - {\rm{E}}[H_{\boldsymbol{\mathcal{\check{A}}}_{\xi}}^{-1}|\xi])\boldsymbol{c} \ge 0
    \end{align}
for any inital design $\xi\in \Xi_{n}$ and all non-zero $\boldsymbol{c}\in\mathbb{R}^{p}$ with equality if and only if $d=1$ or $\gamma=0$.
\end{theorem}
See section \ref{sec:Tech} for proof. Recall the definition of second order efficiency in \eqref{eq:SecondEff}; Theorem \ref{thm:MainRes} states that no \emph{a priori} deterministic design is second order efficient, i.e. \emph{efficiency requires adaptation}. 

As previously stated, the RSLB and the CRLB coincide in normal linear models with constant variance, which corresponds to case where $\gamma=0$ and is excluded. A note on why $d=1$ is excluded. The RRSD algorithm increases the efficiency of an experiment, relative to the deterministic design $\xi$, by adaptively altering the pre-specified allocation weights, $\boldsymbol{w} = (w_{1},\ldots,w_{d})^{T}$, to the pre-specified support points, $\boldsymbol{X} = (\boldsymbol{x}_{1},\ldots,\boldsymbol{x}_{d})^{T}$. If $d=1$ then $\xi$ has only a single support point, with weight $w_{1} = 1$, and $\mathcal{P}_{1}\{r\} = 1$ for all $r$ in Step 4 of Algorithm \ref{alg:RRSD}. When $d=1$ the RRSD results in a design that is equivalent to $\xi$.

\begin{remark} \label{rem:AdaptBias}
Many existing adaptive designs strategies are known to produce biased estimates [\cite{Coad:Ivan:Bias:2001,Bran:Knig:Baue:Esti:2006,Flou:Oron:Bias:2019,Mars:AGen:2021,Lane:Cond:2022}]. This is not expected to be the case for the RRSD. The main intuition for this remark is that the RRSD is ancillary by construction. In other words no information about the parameter $\boldsymbol{\theta}$ is used in the design and is not expected to induce the additional bias observed in existing adaptive designs.
\end{remark}

\subsection{Extension to Randomized Strategies} \label{subsec:MainRandom}

Theorem \ref{thm:MainRes} restricts the initial design to be from the set of \emph{a priori} deterministic designs. The extension of Algorithm \ref{alg:RRSD} to \emph{a priori} random designs is provided in Section \ref{subsec:Random}. The important aspect of the extension is to first, randomly generate a design $\xi$ from $\pi$; then adapt by initializing the RRSD with this observed realization. For the randomize then adapt strategy it can be directly observed that Theorem \ref{thm:MainRes} holds for any given $\xi$ by taking the expectation with respect to the probability measure $\pi$. The following corollary formalizes this statement and extends Theorem \ref{thm:MainRes} to randomized design strategies.
\begin{corollary}
Under the conditions of Theorem \ref{thm:MainRes} for any probability measure such that $\pi(\xi|\boldsymbol{\mathcal{Y}}) = \pi(\xi)$ with probability space $(\mathscr{X},\Xi_{n},\pi)$ 
\begin{align}
    \lim_{n\rightarrow\infty} n^{2}\boldsymbol{c}^{T} \left( {\rm{E}}[ H_{\boldsymbol{\mathcal{A}}_{\pi}}^{-1}] - {\rm{E}}[H_{\boldsymbol{\mathcal{\check{A}}}_{\pi}}^{-1}] \right) \boldsymbol{c} \ge 0
\end{align}
for all non-zero $\boldsymbol{c}\in\mathbb{R}^{p}$ with equality if and only if $d=1$ or $\gamma^{2}=0$.
\end{corollary}
As stated previously, the RRSD is only guaranteed to be better than the \emph{a priori} design used to initialize it. For example, a RRSD initialized by a randomized strategy is only guaranteed to be more efficient than the corresponding \emph{a priori} randomized strategy. It is not guaranteed to be more efficient than a similar \emph{a priori} deterministic design. Section \ref{subsubsec:Random} provides an illustrative example of this discussion.



\section{Deterministic Relevant Subset Design} \label{sec:DRSD}

The RRSD is, in a sense, a proof of concept design. It was constructed to provide a theoretically tractable framework to prove that an adaptive design always exists that is more efficient than any \emph{a priori} design. While the RRSD represents a convenient solution; it is likely inefficient compared to \emph{adaptive} deterministic alternatives. An adaptive deterministic design is distinguished from an \emph{a priori} deterministic design in that, for the former, given the data from the first $r$ runs the design of run $r+1$ is completely determined; whereas, in the latter the design is completely determined in advance of the experiment.   Randomization has many known benefits; but, efficiency is not necessarily one of them. Even in cases where random strategies outperform deterministic ones, e.g. the example in Section \ref{subsubsubsec:Minmax}, the randomization should be implemented before the adaptive design is initialized. It is likely that deterministic alternatives to the RRSD will lead to greater efficiency.

Below a one-step ahead deterministic relevant subset design is proposed. In a one-step ahead setting each run, $r$, corresponds to a single observations (i.e. $r = j$) and the number of runs is equal to the sample size.  
\begin{algorithm} \label{alg:DRSD} Deterministic Relevant Subset Design (DRSD)
\begin{enumerate}
    \item{Repeat steps 1-3 of Algorithm \ref{alg:RRSD}.}
    \item{For $r=n\{1\}+1,\ldots,n$ let $\mathcal{P}_{i}\{r+1\} = w_{i} -  u_{\boldsymbol{\mathcal{A}}_{i}\{1;r\}}/\mathcal{N}\{r+1\}$.}
     \item{Find $i\{r+1\} = \arg\max_{i=1,\ldots,d} \mathcal{P}_{i}\{r+1\}$.
     \item{Allocate the $\{r+1\}$th observation to the support point to $x_{i\{r+1\} }$}.}
\end{enumerate}
\end{algorithm}

Let $\mathring{\xi} = [\mathring{x}(1),\ldots,\mathring{x}(n)]^{T}$ represent the design observed following the DRSD with initial design $\xi$, where $\mathring{x}(j)$ is the support point for observation $j$ assigned by the DRSD algorithm. Additionally, $\boldsymbol{\mathcal{\mathring{A}}}_{\xi}$ and $\tilde{\boldsymbol{\theta}}_{\mathring{\xi}}$ are an ancillary variable and estimate of $\boldsymbol{\theta}$ following the DRSD initialized with $\xi$.

Intuitively, selecting design points one at a time will further increase efficiency. It is not trivial to extend the main results to the DRSD  and such results are not provided in the current work. Instead the justification for the DRSD will be demonstrated in a series of simulation studies in the next section. 

\section{Simulation Study} \label{sec:Examples}
In this section the efficiency of the RRSD and DRSD are compared to \emph{a priori} alternatives in a variety of settings. 

\subsection{Difference in the Means}

Return to the difference in the means example considered in Sections \ref{sec:Intro} and \ref{subsubsec:GND} and recall that the balanced design, $\xi_{\rho}$, is the most efficient fixed design. To demonstrate the RRSD and DRSD algorithms, let $\alpha = \beta = 1/4$, $\eta_{1} = \eta_{2} = 1$ and $n\{1\}=4$. Following Step 1 of Algorithm 1 the adaptive procedures are initialized by placing 2 observations on each treatment. Suppose $\boldsymbol{a}_{1}\{1\} = \{0.8,1.1\}$ and $\boldsymbol{a}_{2}\{1\} = \{1.8,2.2\}$ are observed. Based on these values $u_{\boldsymbol{\mathcal{A}}_{1}\{1\}} = -1.05$, $u_{\boldsymbol{\mathcal{A}}_{2}\{1\}} = 1.05$, $\mathcal{N}\{2\} = 3$, $\mathcal{P}_{1}\{2\} = 0.85$, $\mathcal{P}_{2}\{2\} = 0.15$ and run 2 of the RRSD contains three observation which are assigned to treatment 1 with probability 0.85 and to treatment 2 with probability 0.15. Recall, for the DRSD $r = j$ and it can be seen that $i\{5\} = \arg\max_{i} \mathcal{P}_{i}\{5\} = 1$, i.e. the 5th observation is assigned to treatment 1. 

Recall, $\check{\xi}_{\rho}$ and $\mathring{\xi}_{\rho}$ denote the RRSD and DRSD  initialized by $\xi_{\rho}$. Additionally $\boldsymbol{\mathcal{A}}_{\xi_{\rho}}$, $\boldsymbol{\mathcal{\check{A}}}_{\xi_{\rho}}$ and $\boldsymbol{\mathcal{\mathring{A}}}_{\xi_{\rho}}$ represent an ancillary random variable from $\xi_{\rho}$, $\check{\xi}_{\rho}$ and the $\mathring{\xi}_{\rho}$, respectively.

To compare the performance of the \emph{a prior} deterministic design, $\xi_{\rho}$; the RRSD, $\check{\xi}_{\rho}$; and the DRSD, $\mathring{\xi}_{\rho}$, their efficiencies, defined as 
\begin{align}
    \mbox{LB-EFF-}\xi_{\rho} &= {\rm{E}}[H_{\boldsymbol{\mathcal{A}}_{\xi_{\rho}}}^{-1}|\xi_{\rho}]/\boldsymbol{F}_{\xi_{\rho}}^{-1}, \enspace \mbox{LB-EFF-}\check{\xi}_{\rho} = {\rm{E}}[H_{\boldsymbol{\mathcal{\check{A}}}_{\xi_{\rho}}}^{-1}|\xi_{\rho}]/\boldsymbol{F}_{\xi_{\rho}}^{-1} \\
    \mbox{LB-EFF-}\mathring{\xi}_{\rho} &= {\rm{E}}[H_{\boldsymbol{\mathcal{\mathring{A}}}_{\xi_{\rho}}}^{-1}|\xi_{\rho}]/\boldsymbol{F}_{\xi_{\rho}}^{-1},
\end{align}
are reported. Efficiency is defined relative to the CRLB and is bounded above by 1; the larger the value the more efficient the design. 

\begin{figure}
\centering
\begin{subfigure}{0.495\linewidth}
        \includegraphics[width=\textwidth]{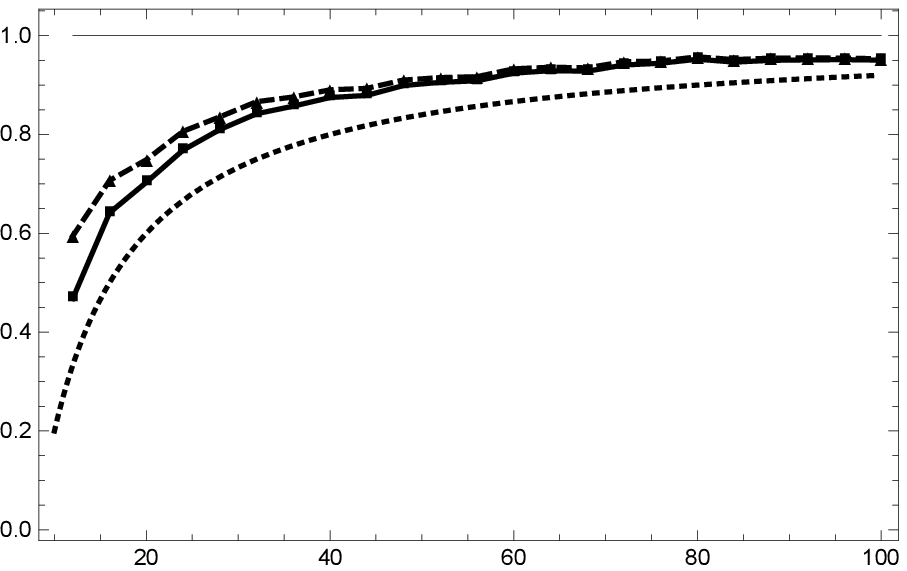}
\end{subfigure} 
\begin{subfigure}{0.495\linewidth}
        \includegraphics[width=\textwidth]{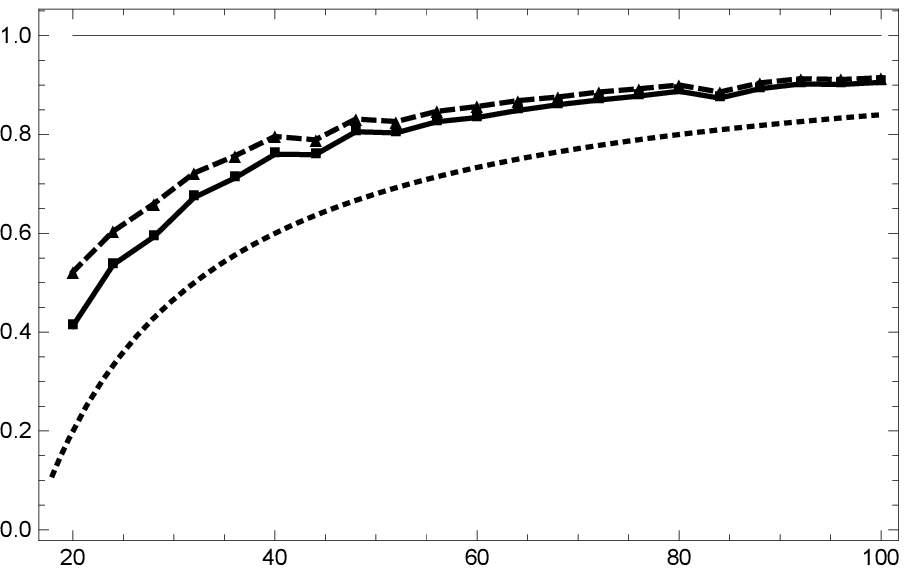}
\end{subfigure}
\caption{LB-EFF-$\check{\xi}_{\rho}$ (solid), LB-EFF-$\mathring{\xi}_{\rho}$ (dashed) and $\mbox{LB-EFF-}\xi_{\rho}$ (dotted) on the $y$-axis for $\alpha = \beta = 1/4$ (left) and $\alpha = \beta = 1/8$ (right) for a variety of sample sizes $n$ on the $x$-axis}\label{fig:Normal}
\end{figure}

Figure \ref{fig:Normal} plots $\mbox{LB-EFF-}\xi_{\rho}$ (solid), $\mbox{LB-EFF-}\check{\xi}_{\rho}$ (dashed) and $\mbox{LB-EFF-}\mathring{\xi}_{\rho}$ (dotted) on the $y$-axis for $\alpha = \beta = 1/4$ (left) and $\alpha = \beta = 1/8$ (right) as a function of the sample size, $n$, on the $x$-axis. The thin horizontal line at one represents the CRLB. $\mbox{LB-EFF-}\xi_{\rho}$ is based on the exact results of Section \ref{subsubsec:GND}; whereas, $\mbox{LB-EFF-}\check{\xi}_{\rho}$ and $\mbox{LB-EFF-}\mathring{\xi}_{\rho}$ are obtained from a simulation with 2000 iterations. As predicted by Theorem \ref{thm:MainRes} the solid curve corresponding to the RRSD is uniformly greater than the dotted curve corresponding to the \emph{a prior} deterministic design. Additionally, as predicted in Section \ref{sec:DRSD} the dashed line corresponding to the DRSD is uniformly more efficient than the RRSD. The gain in efficiency can be quite significant, for example when $\alpha = \beta = 1/8$ and $n = 36$ the efficiency of the DRSD, the RRSD and the \emph{a prior} deterministic design are 0.76, 0.71 and 0.56, respectively. 

LB-EFF measures the efficiency with respect to the RSLB; intuitively this implies that for any estimate of the difference in the means the adaptive designs are expected to be more efficient than the corresponding \emph{a priori} alternative. One convenient feature of the current setting is that the variance of the MLE and the RSLB are equivalent. This immediately implies that RRSD and DRSD reduce the variance of the MLE in the exact same manner that the RSLB was reduced.

\subsection{Optimal Design}
A primary objective in optimal design is to minimize $\boldsymbol{F}_{\xi}$ with respect to a convex optimality criterion, denoted $\Psi(\cdot)$. Formally, an \emph{a priori} deterministic design $\xi_{\Psi}$ is $\Psi$-optimal if
\begin{align} \label{eq:Opt}
\xi_{\Psi} = \arg \min_{\xi \in \Xi_{n}} \Psi(\boldsymbol{F}_{\xi}),
\end{align}
The optimal design has long been considered a benchmark of efficiency, see \cite{Kief:Opti:1959,Kief:Wolf:Opti:1959,Kief:Wolf:TheE:1960} for early foundational works; see \cite{Puke:Opti:2006} and \cite{Atki:Done:Tobi:opti:2007} for reference texts. The $D$-criterion, $\Psi(\boldsymbol{F}_{\xi}) = |\boldsymbol{F}_{\xi}|^{-1/p}$, and the $A$-criterion defined, for non-singular $\boldsymbol{F}_{\xi}$, as $\Psi(\boldsymbol{F}_{\xi}) = \mathrm{Tr}(\boldsymbol{F}_{\xi}^{-1})$ are two commonly used criteria. The definition of efficiency in \eqref{eq:Opt} is narrower than considered in the current work, where efficiency is in Loewner order. 

The significance of Theorem \ref{thm:MainRes} is that implementing a RRSD improves any \emph{a prior} design in Loewner order. This implies that the RRSD is more efficient with respect to most of the so called alphabetic optimally criteria $A$, $D$, $c$, $E$, $G$, $L$, etc., see \cite{Atki:Done:Tobi:opti:2007} for a description of commonly used criteria. For example, suppose a RRSD is initialized with the $D$-optimal design, $\xi_{D}$, not only is the RRSD more efficient with respect to the $D$ criterion, it is also more efficient with respect to all of the other aforementioned alphabetic criteria.

\subsubsection{G-Optimal (Minimax) Design} \label{subsubsec:Random}
A nice illustration of the improvement in efficiency offered by the adaptive approaches is the minimax example outlined in Section \ref{subsubsubsec:Minmax}. The minimax \emph{a priori} deterministic design corresponds to the $G$-optimal design with $\Psi(\boldsymbol{F}_{\xi}) = \max_{x\in\mathscr{X}} \boldsymbol{{\rm{f}}}^{T}(x)\boldsymbol{F}_{\xi}^{-1}\boldsymbol{{\rm{f}}}(x)$, $\mathscr{X} = [-1,1]$ and $\boldsymbol{{\rm{f}}}(x)=(1,x,x^{2})^{T}$. for this simulation study, it is assumed the errors follow a generalized normal distribution (GND), with $\boldsymbol{\theta} = (1,1,1)^{T}$, $\zeta = 10$ and $\tau = 1$, as described in Section \ref{subsubsec:GND}. Recall, that if $n$ is divisible by 3 then $\xi_{G}=\pi_{G}$ place equal weight on the points -1, 0 and 1. For this reason only sample sizes such that $n \;\mathrm{mod}\; 3 = 1$ are examined. The adaptive designs are initialized with $n\{1\} = 6$ with 2 observations placed on each point (-1, 0 and 1) in the first run.

This simulation study considers four designs; the two \emph{a priori} designs, $\xi_{G}$ and $\pi_{G}$; and the two corresponding DRSDs initialized by $\xi_{G}$ and $\pi_{G}$, denoted $\mathring{\xi}_{G}$ and DRSD$_{\pi_{G}}$, respectively. As was the case in the difference in the means example, the RRSD was slightly less efficient than the DRSD; for clarity of exposition the RRSD results are not included. 

In the difference in the means example the ${\rm{Var}}[\boldsymbol{\hat{\theta}}|\xi] = {\rm{E}}[H_{\boldsymbol{\mathcal{A}}}^{-1}|\xi]$. For GND errors this is not the case; in this section the efficiency of the RSLB and the variance of the MLE are examined with respect to certain convex optimality criteria. Specifically, the $\Psi$-efficiency of the RSLB and variance of the MLE for the \emph{a priori} deterministic design $\xi_{G}$  defined as
\begin{align}
    \mbox{LB-EFF$_{\Psi}$-}\xi_{G} &= \frac{{\Psi}\{{\rm{E}}[H_{\boldsymbol{\mathcal{A}}_{\xi_{G}}}^{-1}|\xi_{G}]\}}{{\Psi}\{\boldsymbol{F}_{\xi_{\Psi}}^{-1}\}} \enspace \mbox{and} \enspace \mbox{Var-EFF$_{\Psi}$-}\xi_{G} = \frac{{\Psi}\{{\rm{Var}}[\hat{\boldsymbol{\theta}}_{\xi_{G}}|\xi_{G}]\}}{{\Psi}\{\boldsymbol{F}_{\xi_{\Psi}}^{-1} \}},
\end{align}
respectively, are reported; recall that $\hat{\boldsymbol{\theta}}_{\xi}$ is the MLE following an experiment with design $\xi$. Similarly, $\mbox{LB-EFF$_{\Psi}$-}\pi_{G}$ and $\mbox{Var-EFF$_{\Psi}$-}\pi_{G}$ is the $\Psi$-efficiency of the \emph{a priori} random minimax design; $\mbox{LB-EFF$_{\Psi}$-}\mathring{\xi}_{G}$ and $\mbox{Var-EFF$_{\Psi}$-}\mathring{\xi}_{G}$ is the $\Psi$-efficiency of a DRSD initialized with $\xi_{G}$; and $\mbox{LB-EFF$_{\Psi}$-}\mathring{\pi}_{G}$; and $\mbox{Var-EFF$_{\Psi}$-}\mathring{\pi}_{G}$ is the $\Psi$-efficiency of a DRSD initialized with $\pi_{G}$. All of the above measures are defined with respect to the CRLB corresponding to the $\Psi$-optimal design, $\xi_{\Psi}$, and are bounded above by 1.

\begin{figure}
\centering
\begin{subfigure}{0.5\linewidth}
       \includegraphics[width=\textwidth]{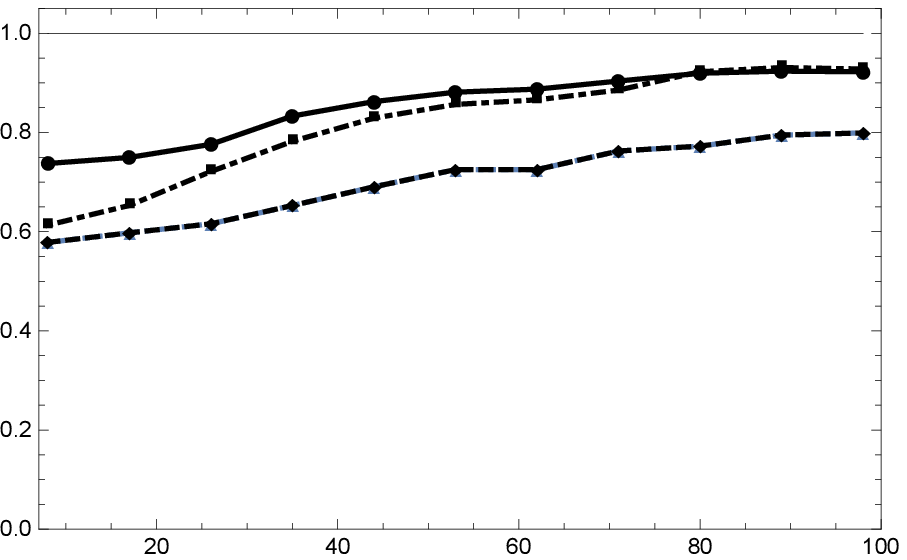}
       \includegraphics[width=\textwidth]{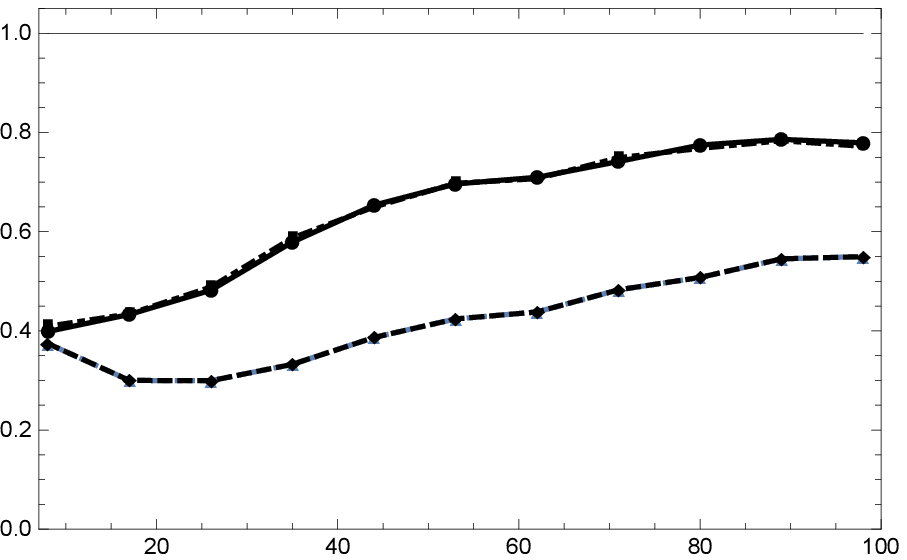}
       \caption{LB-EFF} \label{subfig:D_AOD_MM}
\end{subfigure}\hfill
\begin{subfigure}{0.5\linewidth}
       \includegraphics[width=\textwidth]{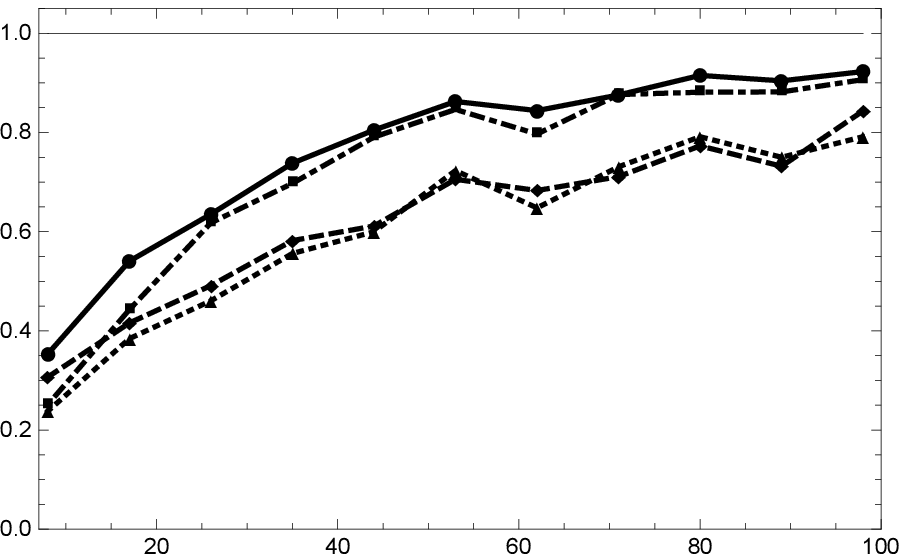}
       \includegraphics[width=\textwidth]{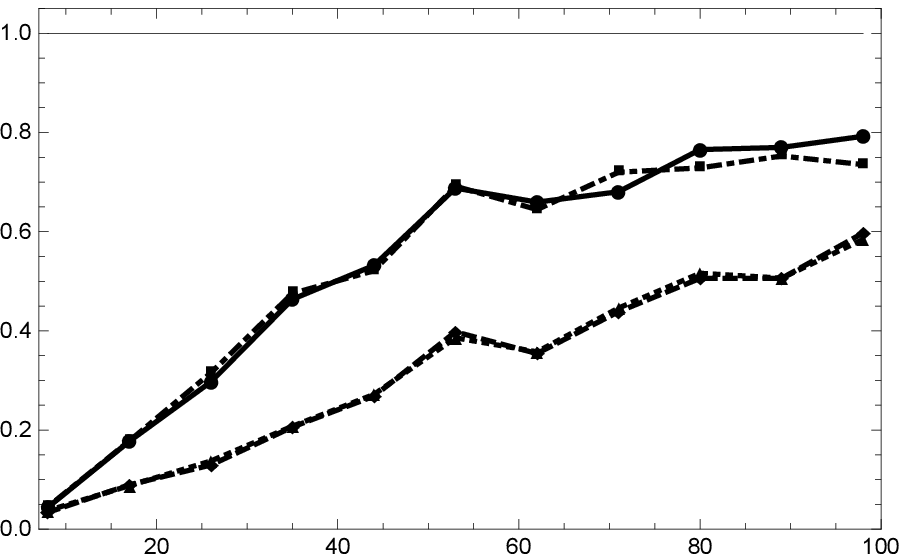}
       \caption{Var-EFF} \label{subfig:D_AOD_TPC}
\end{subfigure}\hfill
\caption{$\mbox{LB-EFF$_{\Psi}$-}\xi_{G}$ (dotted), $\mbox{LB-EFF$_{\Psi}$-}\pi_{G}$ (dashed), $\mbox{LB-EFF$_{\Psi}$-}\mathring{\xi}_{G}$ (dot-dashed) and $\mbox{LB-EFF$_{\Psi}$-}\mathring{\pi}_{G}$ (solid) for $\Psi$ equal to the $G$ (top left) and $D$ (bottom left). $\mbox{Var-EFF$_{\Psi}$-}\xi_{G}$ (dotted), $\mbox{Var-EFF$_{\Psi}$-}\pi_{G}$ (dashed), $\mbox{Var-EFF$_{\Psi}$-}\mathring{\xi}_{G}$ (dot-dashed) and $\mbox{Var-EFF$_{\Psi}$-}\mathring{\pi}_{G}$ (solid) for $\Psi$ equal to the $G$ (top right) and $D$ (bottom right). Sample sizes $n= 8$ to 100 such that $n\mod 3 = 1$ were considered.}\label{fig:GGD}
\end{figure}

The top left panel of Figure \ref{fig:GGD} plots $\mbox{LB-EFF$_{G}$-}\xi_{G}$ (dotted), $\mbox{LB-EFF$_{G}$-}\pi_{G}$ (dashed), $\mbox{LB-EFF$_{G}$-}\mathring{\xi}_{G}$ (dot-dashed) and $\mbox{LB-EFF$_{G}$-}\mathring{\pi}_{G}$ (solid), obtained from 2000 iterations, on the $y$ axis for various sample sizes $n$ such that $n \;\mathrm{mod}\; 3 = 1$ on the $x$ axis. The thin horizontal line at one corresponds to the CRLB. As expected from Theorem \ref{thm:MainRes}, the DRSDs, $\mathring{\pi}_{G}$ and $\mathring{\pi}_{G}$, are more efficient than their respective \emph{a priori} initial designs, $\xi_{G}$ and $\pi_{G}$. An additional feature is that the RSLB for the two \emph{a priori} designs coincide; whereas, $\mathring{\pi}_{G}$ is nearly uniformly more efficient than $\mathring{\pi}_{G}$. Interestingly, this indicates that a randomized strategy does not in itself improve the RSLB; however, initializing the DRSD with $\pi_{G}$, in place of $\xi_{G}$ can improve the RSLB. Perhaps random strategies are more useful in an adaptive setting than in an \emph{a priori} setting. 

The top right panel of Figure \ref{fig:GGD} plots $\mbox{Var-EFF$_{G}$-}\xi_{G}$ (dotted), $\mbox{Var-EFF$_{G}$-}\pi_{G}$ (dashed), $\mbox{Var-EFF$_{G}$-}\mathring{\xi}_{G}$ (dot-dashed) and $\mbox{Var-EFF$_{G}$-}\mathring{\pi}_{G}$ (solid) on the $y$ axis for the same series of simulations presented in the top left panel. The solid curve, which corresponds the DRSD$_{\pi_{G}}$, is uniformly greater than the dashed curve, which corresponds to $\pi_{G}$. Similarly, the dot-dashed curve is uniformly greater than the dotted curve which confirms that the $\mathring{\xi}_{G}$ is uniformly more efficient than $\xi_{G}$. These statements support the claim that the adaptive procedures are more efficient than their initial design.  For the sample size $n = 8$, the efficiency of the \emph{a priori} random design $\pi_{G}$ is 0.31 is greater than the $\mathring{\pi}_{G}$ efficiency of 0.25. This illustrates the point that the proposed adaptive strategies are only guaranteed to improve the initial design. For comparison the efficiency of $\xi_{G}$ and $\mathring{\pi}_{G}$ are 0.24 and 0.35, respectively.  Note, for all other sample sizes examined $\mathring{\pi}_{G}$ was more efficient than $\pi_{G}$, which indicates that the benefit of the \emph{a priori} random strategy disappears quickly as $n$ increases.

The preceding discussion concerns the efficiency of the various designs with respect to the $G$-criterion. One important aspect of Theorem \ref{thm:MainRes} is that the improvement is in Loewner order. To illustrate, the bottom row of Figure \ref{fig:GGD} plots the same series of simulations and efficiency measures as the top row for $\Psi$ equal to the $D$-criterion. For the $D$-criterion the efficiency of the RSLB (bottom left) and the variance of the MLE (bottom right) for $\xi_{G}$ and $\pi_{G}$ are in perfect correspondence. This equality occurs because the $G$-optimal, $\pi_{G}$, offers no guarantee for the $D$-, or any other, optimality criteria. The solid and dot-dashed curves representing the $\mathring{\xi}_{G}$ and DRSD$_{\pi_{G}}$ offer a nice illustration of the global efficiency of proposed adaptive designs. The two adaptive procedures nearly perfectly coincide and are both uniformly more efficient than the \emph{a priori} designs $\xi_{G}$ and $\pi_{G}$ with respect to the $D$-criterion. Repeating the analysis for any of the commonly used alphabetic criteria produces a similar result. This is a direct result of the improvement in Loewner order.

\subsection{Factorial Design}

Consider an experiment with two treatment factors $\boldsymbol{x} = (x_{1},x_{2})^{T}$, design space $\mathscr{X} =[0,1]^{2}$, $\boldsymbol{{\rm{f}}}^{T}(\boldsymbol{x}) = (1, x_{1}, x_{2}, x_{1}x_{2})^{T}$ and $\boldsymbol{\theta} = (1,1,1,1)^{T}$. The factorial design, $\xi_{F} = \{ (0,0), (0,1), (1,0), (1,1)\}$, places equal weight on each point and allows for the efficient estimation of the main and interaction effects. For $n\{1\} = 8$ the first run places 2 observations on each point in $\xi_{F}$. It is assumed that the errors are Cauchy distributed, with $\tau=1$, as described in Section \ref{subsubsec:Cauchy}. 

The top row of Table \ref{tab:FD} reports the RSLB, $n{\rm{E}}[H_{\boldsymbol{\mathcal{A}}}^{-1}|\xi_{F}]$, for $\boldsymbol{\mathcal{A}}$ equal to $\boldsymbol{\mathcal{A}}_{\xi_{F}}$ (left), $\boldsymbol{\mathcal{\check{A}}}_{\xi_{F}}$ (middle) and $\boldsymbol{\mathcal{\mathring{A}}}_{\xi_{F}}$ (right) from 2000 iterations for the experiment described above with sample size $n=60$. The second row of Table \ref{tab:FD} reports the variance of the MLE, $n{\rm{Var}}[\tilde{\boldsymbol{\theta}}_{\xi}|\xi_{F}]$, for $\xi$ equal to $\xi_{F}$ (left), $\check{\xi}_{F}$ (middle) and $\mathring{\xi}_{F}$ (right) for the same scenario as in the top row. The bottom row of Table \ref{tab:FD} reports the CRLB, $n\boldsymbol{F}_{\xi_{F}}^{-1}$, for reference.

Recall the discussion following Proposition \ref{prop:LBexp} which stated that the objective of the adaptive procedures was to reduce the difference between the RSLB and CRLB. The top row of Table \ref{tab:FD} confirms that is objective is accomplished; the RSLB corresponding to $\check{\xi}_{F}$ (column 2) and $\mathring{\xi}_{F}$ (column 3) are markedly closer to $\boldsymbol{F}_{\xi_{F}}^{-1}$ than the RSLB corresponding to $\xi_{F}$ (column 1). The RSLB is a lower bound for all estimates of $\boldsymbol{\theta}$; as such it is expected that the adaptive procedures will reduce the variance of estimates of $\boldsymbol{\theta}$. For the MLE this benefit is confirmed in the second row of Table \ref{tab:FD} where it can be seen that the variance of the MLE obtained following the $\check{\xi}_{F}$ (column 2) and $\mathring{\xi}_{F}$ (column 3) are markedly closer to $\boldsymbol{F}_{\xi_{F}}^{-1}$ than the variance of the MLE from an experiment with design $\xi_{F}$ (column 1). 

\begin{table}
\setlength\arraycolsep{2pt} 
\setlength\tabcolsep{0pt}

\caption{The RSLBs (top row), $n{\rm{E}}[H_{\boldsymbol{\mathcal{A}}_{\xi_{F}}}^{-1}|\xi_{F}]$ (left)
, $n{\rm{E}}[H_{\boldsymbol{\mathcal{\check{A}}}_{\xi_{F}}}^{-1}|\xi_{F}]$ (middle) and $n{\rm{E}}[H_{\boldsymbol{\mathcal{\mathring{A}}}_{\xi_{F}}}^{-1}|\xi_{F}]$ (right); and the variance of the MLEs (second row), $n{\rm{Var}}[\hat{\boldsymbol{\theta}}_{\xi_{F}}|\xi_{F}]$ (left),  $n{\rm{Var}}[\hat{\boldsymbol{\theta}}_{\check{\xi}_{F}}|\xi_{F}]$ (middle)
and $n{\rm{Var}}[\hat{\boldsymbol{\theta}}_{\mathring{\xi}_{F}}|\xi_{F}]$ (right) for a factorial experiment from 2000 iterations with sample size $n=60$. The CRLB (bottom row), $n\boldsymbol{F}_{\xi_{F}}^{-1}$, is provided for reference.}\label{tab:FD}

\begin{tabular}{ccc}
\toprule 
$n{\rm{E}}[H_{\boldsymbol{\mathcal{A}}_{\xi_{F}}}^{-1}|\xi_{F}]$
& $n{\rm{E}}[H_{\boldsymbol{\mathcal{\check{A}}}_{\xi_{F}}}^{-1}|\xi_{F}]$
& $n{\rm{E}}[H_{\boldsymbol{\mathcal{\mathring{A}}}_{\xi_{F}}}^{-1}|\xi_{F}]$ \\ 
\midrule 
$\left( \begin{array}{cccc} 
9.2 & -9.2 & -9.2 & 9.2 \\ 
-9.2 & 18.3 & 9.2 & -18.3 \\ 
-9.2 & 9.2 & 18.5 & -18.5 \\
9.3 & -18.3 & -18.5 & 36.9
\end{array}\right)$ &
$\left( \begin{array}{cccc} 
8.8 & -8.8 & -8.8 & 8.8 \\ 
-8.8 & 17.7 & 8.8 & -17.7 \\ 
-8.8 & 8.8 & -17.7 & -17.7 \\
8.8 & -17.7 & -17.7 & 35.4
\end{array}\right)$
  & 
$\left( \begin{array}{cccc} 
8.3 & -8.3 & -8.3 & 8.3 \\ 
-8.3 & 16.6 & 8.3 & -16.6 \\ 
-8.3 & 8.3 & 16.6 & -16.6 \\
8.3 & -16.6 & -16.6 & 33.3
\end{array}\right)$
\\  \midrule 
$n{\rm{Var}}[\hat{\boldsymbol{\theta}}_{\xi_{F}}|\xi_{F}]$
& $n{\rm{Var}}[\hat{\boldsymbol{\theta}}_{\check{\xi}_{F}}|\xi_{F}]$
& $n{\rm{Var}}[\hat{\boldsymbol{\theta}}_{\mathring{\xi}_{F}}|\xi_{F}]$ \\ 
\midrule 
$\left( \begin{array}{cccc} 
10.1 & -10.2 & -10.2 & 10.6 \\ 
-10.2 & 20.9 & 10.4 & -21.2 \\ 
-10.2 & 10.4 & 19.7 & -19.8 \\
10.6 & -21.1 & -19.9 & 40.7
\end{array}\right)$ &
$\left( \begin{array}{cccc} 
9.4 & -9.3 & -9.4 & 9.6 \\ 
-9.3 & 18.7 & 9.4 & -18.8 \\ 
-9.4 & 9.4 & 18.8 & -19.3 \\
9.6 & -18.8 & -19.3 & 38.2
\end{array}\right)$
  & 
$\left( \begin{array}{cccc} 
8.5 & -8.4 & -8.8 & 8.9 \\ 
-8.4 & 17.0 & 8.8 & -17.3 \\ 
-8.8 & 8.8 & 17.7 & -17.6 \\
8.9 & -17.3 & -17.6 & 34.1
\end{array}\right)$
\\  \midrule 
& $n\boldsymbol{F}_{\xi_{F}}^{-1}$
&  \\ 
\midrule 
 &
$\left( \begin{array}{cccc} 
8.0 & -8.0 & -8.0 & 8.0 \\ 
-8.0 & 16.0 & 8.0 & -16.0 \\ 
-8.0 & 8.0 & 16.0 & -16.0 \\
8.0 & -16.0 & -16.0 & 32.0
\end{array}\right)$
  & 
\\
\bottomrule       
\end{tabular}

\end{table}


\section{Discussion} \label{sec:Discuss}

In this work two adaptive designs were developed, the randomized relevant subset design (RRSD) and the deterministic relevant subset design (DRSD), that are more efficient than any corresponding $\emph{a priori}$ design, with respect to Loewner order. An $\emph{a priori}$ design is any design that is independent of the observed data and can be determined completely in advance of the experiment. This improvement was accomplished by first, deriving a relevant subset lower bound (RSLB) that represents a tighter alternative to the Cram\'{e}r-Rao lower bound (CRLB). And second, by developing an adaptive algorithm that incorporate the relevant subset into the design of the experiment with the expressed intent of reducing the RSLB. 

The main theoretical results in this work are restricted to asymptotic efficiency. A simulation study demonstrated the finite sample benefits of the proposed adaptive designs. Both the RRSD and DRSD were uniformly more efficient than the corresponding $\emph{a priori}$ alternatives for a variety of examples. 

The proposed adaptive designs are initialize by an \emph{a priori} design.  The adaptive designs are only guaranteed to be more efficient than the initial $\emph{a priori}$ design. It is critical for the overall performance that the RRSD and DRSD are initialized using efficient \emph{a priori} designs. Efficient \emph{a priori} design strategies are critical to the success of the RRSD and DRSD. 


\section{Technical Details} \label{sec:Tech}
In this section the supporting arguments for the main theoretical results are presented. Technical details for supporting lemmas are provided in the supplemental materials.

\subsection{Proof of Lemma \ref{lem:TwoStage}} \label{sec:VarBound}

The check notation is used to indicate an object corresponds to the RRSD, e.g. $\boldsymbol{\check{\mathcal{Y}}}$, $\boldsymbol{\mathcal{\check{A}}}$ and $\mathcal{\check{S}}_{i}$. This notation is introduced for the RRSD in the paragraph following Algorithm \ref{alg:RRSD}. The following lemma is required for the validity of equation \eqref{eq:uai2}. 
\begin{lemma} \label{lem:Ameasurable}
$\boldsymbol{\mathcal{\check{A}}}\{1\}$ is $\boldsymbol{\mathcal{\check{A}}}\{1;2\}$ measurable.
\end{lemma}
The second term in the right hand side of \eqref{eq:VarUA12} can be written as
\begin{align} \label{eq:EVarUA}
    {\rm{E}}[{\rm{Var}}[u_{\boldsymbol{\mathcal{\check{A}}}_{i}\{1;2\}}|\boldsymbol{\mathcal{\check{A}}}_{i}\{1\}]] = {\rm{E}}[{\rm{Var}}[{\rm{E}}[\mathcal{\check{S}}_{i}\{1;2\}|\boldsymbol{\mathcal{\check{A}}}_{i}\{1;2\}]|\boldsymbol{\mathcal{\check{A}}}_{i}\{1\}]]
\end{align}
byt the definition of $u_{\boldsymbol{\mathcal{\check{A}}}_{i}}$. 
\begin{lemma} \label{lem:VarS}
Under the current conditions 
\begin{align}
    {\rm{Var}}[{\rm{E}}[\mathcal{\check{S}}_{i}\{1;2\}|\boldsymbol{\mathcal{\check{A}}}_{i}\{1;2\}]|\boldsymbol{\mathcal{\check{A}}}_{i}\{1\}] = {\rm{Var}}[{\rm{E}}[\mathcal{\check{S}}_{i}\{2\}|\boldsymbol{\mathcal{\check{A}}}_{i}\{1;2\}]|\boldsymbol{\mathcal{\check{A}}}_{i}\{1\}] + O(1)
\end{align}
\end{lemma}
Lemma \ref{lem:VarS} implies that the right hand side of \eqref{eq:EVarUA} is
\begin{align}
    {\rm{E}}[{\rm{Var}}[{\rm{E}}[\mathcal{\check{S}}_{i}\{1;2\}|\boldsymbol{\mathcal{\check{A}}}_{i}\{1;2\}]|\boldsymbol{\mathcal{\check{A}}}_{i}\{1\}]] = {\rm{E}}[{\rm{Var}}[{\rm{E}}[\mathcal{\check{S}}_{i}\{2\}|\boldsymbol{\mathcal{\check{A}}}_{i}\{1;2\}]|\boldsymbol{\mathcal{\check{A}}}_{i}\{1\}]] + O(1).
\end{align}
The bulk of the remainder of the proof derives a bound for the right hand side of the above ignoring the $O(1)$ term. 
\begin{align}
    &{\rm{E}}[{\rm{Var}}[{\rm{E}}[\mathcal{\check{S}}_{i}\{2\}|\boldsymbol{\mathcal{\check{A}}}_{i}\{1;2\}]|\boldsymbol{\mathcal{\check{A}}}_{i}\{1\}]] \\
    &\enspace\enspace\le {\rm{E}}[{\rm{Var}}[\mathcal{\check{S}}_{i}\{2\}|\boldsymbol{\mathcal{\check{A}}}_{i}\{1\}]] \\
    &\enspace\enspace= {\rm{E}}\left[\sum_{j=n\{1\}+1}^{\mathcal{N}\{1;2\}}{\rm{Var}}[(\mathcal{T}_{i}(j) - w_{i})s_{\mathcal{\check{Y}}(j)}|\boldsymbol{\mathcal{\check{A}}}_{i}\{1\}]\right] \\
    &\enspace\enspace= {\rm{E}}\left[\sum_{j=n\{1\}+1}^{\mathcal{N}\{1;2\}} {\rm{Var}}[\{\mathcal{\check{T}}_{i}(j) - w_{i}\}{\rm{E}}[s_{\mathcal{\check{Y}}(j)}|\mathcal{\check{T}}_{i}(j)]|\boldsymbol{\mathcal{\check{A}}}_{i}\{1\}]\right. \\
    &\enspace\enspace\enspace\enspace+ \left. {\rm{E}}[\{\mathcal{\check{T}}_{i}(j) - w_{i}\}{\rm{Var}}[s_{\mathcal{\check{Y}}(j)}|\mathcal{\check{T}}_{i}(j)]|\boldsymbol{\mathcal{\check{A}}}_{i}\{1\}] \right] \\
    &\enspace\enspace\le {\rm{E}}\left[\mathcal{N}\{2\}]\right] \\
    &\enspace\enspace=  \left\{{\rm{E}}\left[\max_{i}\{u_{\boldsymbol{\mathcal{\check{A}}}_{i}\{1\}}/w_{i}\}]\right] +1\right\}.
\end{align}
The first line is due to the law of total variance; the second is from the definition of $\mathcal{\check{S}}_{i}\{2\}$, the fact that $(\mathcal{\check{T}}_{i}(j) - w_{i})s_{\mathcal{\check{Y}}(j)}$ $j=n\{1\}+1,\ldots,\mathcal{N}\{1;2\}$ given $\boldsymbol{\mathcal{\check{A}}}_{i}\{1\}$ is a sequence of independent and identically distributed random variables and the fact that $\mathcal{N}\{1;2\}$ is $\boldsymbol{\mathcal{\check{A}}}_{i}\{1\}$ measurable; the third equality is due to the law of total variance; the fourth is obtained by recognising that given $\mathcal{\check{T}}_{i}(j)$ them mean and variance of $s_{\mathcal{\check{Y}}(j)}$ is $\mu$ and 1, respectively, see the proof of Proposition \ref{prop:LBexp} in the supplemental materials; the final line is from the definition of $\mathcal{N}\{2\}$ in \eqref{eq:u2}, where the $+1$ is required to account for the ceiling operator.

Next consider the term
\begin{align}
    {\rm{E}}\left[\max_{i}\{u_{\boldsymbol{\mathcal{\check{A}}}_{i}\{1\}}/w_{i}\}]\right] &\le {\rm{E}}\left[|\max_{i}\{u_{\boldsymbol{\mathcal{\check{A}}}_{i}\{1\}}/w_{i}\}|^{2}]\right]^{1/2} \le {\rm{E}}\left[\sum_{i=1}^{d}(u_{\boldsymbol{\mathcal{\check{A}}}_{i}\{1\}}^{2}/w_{i})^{2}]\right]^{1/2} \\
    &\le w_{i}^{-1}\left\{\sum_{i=1}^{d}{\rm{Var}}\left[u_{\boldsymbol{\mathcal{\check{A}}}_{i}\{1\}}\right]\right\}^{1/2} = O[(n\{1\})^{1/2}].
\end{align}
The first line is due to H{\"o}lder's inequality, the second is due to the fact that a sum of list of squares is greater than the maximum of a list squared, the third is from the \emph{a priori} nature of the first run design, specifically, the independence property of $u_{\boldsymbol{\mathcal{\check{A}}}_{i}\{1\}}$, $i=1,\ldots,d$ and the fact that ${\rm{E}}[u_{\boldsymbol{\mathcal{\check{A}}}_{i}\{1\}}] = 0$; and the final line is again from the property of \emph{a priori} designs where it was previously shown that ${\rm{Var}}[u_{\boldsymbol{\mathcal{\check{A}}}_{i}\{1\}}] = O[(n\{1\})]$. Putting the preceding together yields
\begin{align}
    \lim_{n\{1\}\rightarrow\infty} {\rm{Var}}[u_{\boldsymbol{\mathcal{\check{A}}}\{1;2\}}|\xi]/n\{1\} = 0
\end{align}
as stated. 

\subsection{Proof of Lemma \ref{lem:AdaptRSLB}}

Lemma \ref{supp:lem:ancillary} [supplemental materials] implies that Proposition \ref{prop:CCRLB} extends to the RRSD and taking the expectation implies the first inequality in Propositions \ref{lem:AdaptRSLB}, provided the necessary conditions which are assumed. 

For the second inequality the following lemmas is required.
\begin{lemma} \label{lem:EH} Under the present conditions ${\rm{E}}[h_{\boldsymbol{\mathcal{\check{A}}}_{i}}|\xi] = 0$ for $i=1,\ldots,d$.
\end{lemma}

Now that the expectation of $h_{\boldsymbol{\mathcal{\check{A}}}_{i}}$ has been established the extension of Jensen's inequality given in \cite{Grov:Roth:ANot:1969} implies that ${\rm{E}}[\boldsymbol{H}_{\boldsymbol{\mathcal{\check{A}}}_{\xi}}^{-1}|\xi] \ge F_{\xi}^{-1}$.

\subsection{Proof of Theorem \ref{thm:MainRes}}
Proposition \ref{prop:LBexp} and Lemma \ref{lem:AdaptRSLB} implies that we can write 
\begin{align} \label{eq:Hdiff}
    &\enspace\lim_{n\rightarrow\infty} n^{2}\boldsymbol{c}^{T}({\rm{E}}[\boldsymbol{H}_{\boldsymbol{\mathcal{A}}_{\xi}}^{-1}|\xi] - {\rm{E}}[\boldsymbol{H}_{\boldsymbol{\mathcal{\check{A}}}_{\xi}}^{-1}|\xi])\boldsymbol{c}  \\
    &\enspace\enspace\enspace = \mu \lim_{n\rightarrow\infty} n^{-1} {\rm{Tr}}[\boldsymbol{D} ({\rm{Var}}[\boldsymbol{u}_{\boldsymbol{\mathcal{A}}_{\xi}}|\xi] + {\rm{Var}}[\boldsymbol{v}_{\boldsymbol{\mathcal{A}}_{\xi}}|\xi] - {\rm{Var}}[\boldsymbol{u}_{\boldsymbol{\mathcal{\check{A}}}_{\xi}}|\xi] - {\rm{Var}}[\boldsymbol{v}_{\boldsymbol{\mathcal{\check{A}}}_{\xi}}|\xi] )].
\end{align}
The first lemma establishes that the limiting behavior of ${\rm{Var}}[\boldsymbol{v}_{\boldsymbol{\mathcal{\check{A}}}_{\xi}}|\xi]$.
\begin{lemma} \label{lem:VV} 
Under the present conditions $\lim_{n\rightarrow\infty} n^{-1}{\rm{Var}}[\boldsymbol{v}_{\boldsymbol{\mathcal{\check{A}}}_{\xi}}|\xi]  = \gamma^{2}\boldsymbol{w}\boldsymbol{w}^{T}$.
\end{lemma}
The implications being that the right hand side of \eqref{eq:Hdiff} reduces to ${\rm{Tr}}[\boldsymbol{D} ({\rm{Var}}[\boldsymbol{u}_{\boldsymbol{\mathcal{A}}_{\xi}}|\xi] - {\rm{Var}}[\boldsymbol{u}_{\boldsymbol{\mathcal{\check{A}}}_{\xi}}|\xi]$. The following lemma implies that the term $n^{-1}{\rm{Var}}[\boldsymbol{u}_{\boldsymbol{\mathcal{\check{A}}}_{\xi}}|\xi]$ disappears for large $n$. 
\begin{lemma} \label{lem:VU} Under the present conditions ${\rm{Var}}[\boldsymbol{u}_{\boldsymbol{\mathcal{\check{A}}}_{\xi}}|\xi] = O(n^{1/2})$.
\end{lemma}
From Lemmas \ref{lem:VV} and \ref{lem:VU} we can now state that 
\begin{align} \label{eq:Hdiffred}
   \lim_{n\rightarrow\infty} n^{2}\boldsymbol{c}^{T}({\rm{E}}[\boldsymbol{H}_{\boldsymbol{\mathcal{A}}_{\xi}}^{-1}|\xi] - {\rm{E}}[\boldsymbol{H}_{\boldsymbol{\mathcal{\check{A}}}_{\xi}}^{-1}|\xi])\boldsymbol{c} = \gamma^{2}{\rm{Tr}}[\boldsymbol{D}(\boldsymbol{W} - \boldsymbol{w}\boldsymbol{w}^{T})] \ge 0.
\end{align}
See the proof of Proposition \ref{prop:LBexp} for additional details. The if only if claim is verified by the following lemma.
\begin{lemma} \label{eq:Tr0}
For all $\xi\in\Xi_{n}$, ${\rm{Tr}}[\boldsymbol{D}(\boldsymbol{W} - \boldsymbol{w}\boldsymbol{w}^{T})]\ge 0$ with equality if and only if $d=1$.
\end{lemma}
This concludes the proof of the theorem.

\begin{supplement}

\begin{center}
\textbf{Supplemental Materials for Efficiency Requires Adaptation}
\end{center}
\renewcommand{\theequation}{S\arabic{equation}}

\section{Supplemental Introduction}

All expectations and variances are conditional on the design $\xi$ and thus this is omitted. 

\section{Proof of Proposition \ref{prop:CCRLB}}
First, note that 
\begin{align} \label{supp:eq:Score}
    \frac{\partial}{\partial\boldsymbol{\theta}}\log f(\boldsymbol{y}|\boldsymbol{\theta}) = \sum_{i=1}^{d} \frac{\partial\eta(\boldsymbol{x}_{i},\boldsymbol{\theta})}{\partial\boldsymbol{\theta}} \frac{\partial}{\partial\eta_{i}}\log f(\boldsymbol{y}_{i}|\eta_{i}),
\end{align}
where $\boldsymbol{y}_{i} = \{y(j):t_{i}(j)=1, j=1,\ldots,n\}$. From the ancillarity of $\boldsymbol{a}_{i}$ its distribution is not a function of $\eta_{i}$; therefore,
\begin{align} \label{supp:eq:ScoreAlt}
    \frac{\partial}{\partial\eta_{i}}\log f(\boldsymbol{y}_{i}|\eta_{i}) = \frac{\partial}{\partial\eta_{i}}\left[\log f(\boldsymbol{y}_{i}|\boldsymbol{a}_{i},\eta_{i}) + \log f(\boldsymbol{a}_{i})\right] = \frac{\partial}{\partial\eta_{i}}\log  f(\boldsymbol{y}_{i}|\boldsymbol{a}_{i},\eta_{i}).
\end{align}
Continuing, Condition \ref{cond:RegularExt} (5') ensures that
\begin{align}
    {\rm{E}}\left[\frac{\partial}{\partial\eta_{i}}\log  f(\boldsymbol{\mathcal{Y}}_{i}|\boldsymbol{\mathcal{A}}_{i},\eta_{i}) \mid \boldsymbol{\mathcal{A}}_{i} \right] = 0 \implies {\rm{E}}\left[\frac{\partial}{\partial\boldsymbol{\theta}}\log f(\boldsymbol{\mathcal{Y}}|\boldsymbol{\theta}) \mid \boldsymbol{\mathcal{A}}\right] = 0.
\end{align}
Using the alternate notation of Section \ref{sec:LB} the above implies that ${\rm{E}}[\dot{l}_{\boldsymbol{\mathcal{Y}},\boldsymbol{\mathcal{T}}_{\negmedspace i}}|\boldsymbol{\mathcal{A}}_{i}] = 0$. Equations \eqref{supp:eq:Score} and \eqref{supp:eq:ScoreAlt} imply that $\partial \log f(\boldsymbol{y}|\boldsymbol{\theta})/(\partial\boldsymbol{\theta}) = \partial \log f(\boldsymbol{y}|\boldsymbol{A},\boldsymbol{\theta})/(\partial\boldsymbol{\theta})$; therefore, 
\begin{align}
    {\rm{Cov}} \left[ \tilde{\boldsymbol{\theta}} ,\frac{\partial}{\partial\boldsymbol{\theta}}\log f(\boldsymbol{\mathcal{Y}}|\boldsymbol{\theta}) \mid \boldsymbol{\mathcal{A}} \right] &= {\rm{E}}\left[ \tilde{\boldsymbol{\theta}}\frac{\partial}{\partial\boldsymbol{\theta}}\log f(\boldsymbol{\mathcal{Y}}|\boldsymbol{\mathcal{A}},\boldsymbol{\theta}) \mid \boldsymbol{\mathcal{A}}  \right] \\
    &= \frac{\partial}{\partial\boldsymbol{\theta}} {\rm{E}} [ \tilde{\boldsymbol{\theta}} \mid \boldsymbol{\mathcal{A}} ] = \boldsymbol{0}
\end{align}
from  Condition \ref{cond:RegularExt} (5') and the conditional unbiased assumption. From a final appeal to Condition \ref{cond:RegularExt} (5') it is straightforward to show that
\begin{align} \label{supp:eq:VarScore}
    \quad\enspace {\rm{Var}}\left[\frac{\partial}{\partial\eta_{i}} \log  f(\boldsymbol{\mathcal{Y}}_{i}|\boldsymbol{\mathcal{A}}_{i},\eta_{i}) \mid \boldsymbol{\mathcal{A}}_{i}  \right] = {\rm{E}}\left[\frac{\partial^{2}}{\partial\eta_{i}^{2}} \log  f(\boldsymbol{\mathcal{Y}}_{i}|\boldsymbol{\mathcal{A}}_{i},\eta_{i}) \mid \boldsymbol{\mathcal{A}}_{i} \right] 
    = h_{\boldsymbol{\mathcal{A}}_{i}}.
\end{align}
From \eqref{supp:eq:Score}, \eqref{supp:eq:ScoreAlt}, \eqref{supp:eq:VarScore} and an appeal to the independence of the responses 
\begin{align}
    {\rm{Var}}\left[\frac{\partial}{\partial\boldsymbol{\theta}}\log f(\boldsymbol{y}|\boldsymbol{\theta}) \mid \boldsymbol{\mathcal{A}}\right] = \boldsymbol{H}_{\boldsymbol{\mathcal{A}}}(\boldsymbol{\theta}).
\end{align}
Finally, an appeal to the Cauchy-Schwarz inequality yields ${\rm{Var}}[\tilde{\boldsymbol{\theta}}|\boldsymbol{\mathcal{A}}] \ge \boldsymbol{H}_{\boldsymbol{\mathcal{A}}}^{-1}$ as stated.

\section{Proof of Proposition \ref{prop:RSLB}}
The stated result follows from the below
\begin{align}
     {\rm{Var}}[\tilde{\boldsymbol{\theta}}] &= {\rm{E}}[{\rm{Var}}[\tilde{\boldsymbol{\theta}}|\boldsymbol{\mathcal{A}}]] + {\rm{Var}}[{\rm{E}}[\tilde{\boldsymbol{\theta}}|\boldsymbol{\mathcal{A}}]] \\
     &= {\rm{E}}[{\rm{Var}}[\tilde{\boldsymbol{\theta}}|\boldsymbol{\mathcal{A}}]] \\
     & \ge {\rm{E}}[\boldsymbol{H}_{\boldsymbol{\mathcal{A}}}^{-1}] \\
     &\ge \boldsymbol{F}_{\xi}^{-1}.
\end{align}
where the first line is from the law of total variance; the second is from the unbiased assumption; the third is due to an application of Proposition \ref{prop:CCRLB}; and the final line is obtained from an application of Proposition \ref{prop:CCRLB} and the extension of Jensen's inequality for inverse matrices proven in \cite{Grov:Roth:ANot:1969}.

\section{Proof of Proposition \ref{prop:LBexp}}
Let $\boldsymbol{\mathcal{Z}} = (\mathcal{Z}_{1}, \ldots,\mathcal{Z}_{d})^{T}$, where $\mathcal{Z}_{i} =  n^{-1/2}(h_{\boldsymbol{\mathcal{A}}_{i}} - nw_{i}\mu)$. Under the assumed conditions a Taylor expansion for $\boldsymbol{\mathcal{Z}}$ around its expectation $\boldsymbol{0}$ yields
\begin{align} \label{eq:Hexp}
    n^{2}\boldsymbol{c}^{T}\left({\rm{E}}[H_{\boldsymbol{\mathcal{A}}}^{-1}] - \boldsymbol{F}_{\xi}^{-1}\right)\boldsymbol{c} &= {\rm{E}}[\boldsymbol{Z}^{T}\boldsymbol{D}\boldsymbol{Z}] + o(1).
\end{align}
Let $\mathcal{Z}_{i} =  n^{-1/2}(h_{\boldsymbol{\mathcal{A}}_{i}} - nw_{i}\mu)$ and note that per the assumed conditions $\mathcal{Z}_{i} - \mathcal{Z}_{i} = O(n^{-1/2})$. Therefore, $\mathcal{Z}_{i}$ can be replaced with $\mathcal{Z}_{i}$ in the right hand side of \eqref{eq:Hexp} without impacting the order of approximation. Noting that $\mathcal{Z}_{i} = \mu n^{1/2}(u_{\boldsymbol{\mathcal{A}}_{i}} + v_{\boldsymbol{\mathcal{A}}_{i}})$
\begin{align}
    {\rm{E}}[\boldsymbol{Z}^{*T}\boldsymbol{D}\boldsymbol{Z}] &= {\rm{Tr}}\{\boldsymbol{D}{\rm{Var}}[\boldsymbol{Z}]\} \\
    &= n^{-1}\mu {\rm{Tr}}\{\boldsymbol{D}({\rm{Var}}[\boldsymbol{u}_{\boldsymbol{\mathcal{A}}}] + {\rm{E}}[\boldsymbol{u}_{\boldsymbol{\mathcal{A}}}\boldsymbol{v}_{\boldsymbol{\mathcal{A}}}^{T}] \\
    &\quad\quad+ {\rm{E}}[\boldsymbol{v}_{\boldsymbol{A}}\boldsymbol{u}_{\boldsymbol{\mathcal{A}}}^{T}] + {\rm{Var}}[\boldsymbol{v}_{\boldsymbol{\mathcal{A}}}])\}, \label{supp:eq:ZDZ}
\end{align}
First consider the expectations
\begin{align}
    {\rm{E}}[u_{\boldsymbol{\mathcal{A}}_{i}}] &= {\rm{E}}\left[\sum_{j=1}^{n} \{t_{i}(j) - w_{i}\}i_{\mathcal{Y}(j)}^{(\phi)}\right] = \sum_{j=1}^{n} \{t_{i}(j) - w_{i}\}= 0 \\
    {\rm{E}}[v_{\boldsymbol{\mathcal{A}}_{i}}] &= w_{i}{\rm{E}}\left[\left\{\sum_{j=1}^{n}\{i_{\mathcal{Y}(j)}^{(\phi)}-1\}\right\}\right] = 0.
\end{align}
Let $q_{i} = \sum_{j = 1}^{n}t_{i}(j)q_{y(j)}$, where $q_{y(j)} = [i_{y(j)}^{(\phi)} -1 - \nu_{11}\dot{l}_{y(j)}^{(\phi)}]/\gamma$ has mean 0 and variance 1. Further, let $\boldsymbol{Q} = (q_{1},\ldots,q_{d})^{T}$. 
\begin{lemma} \label{supp:lem:EHA}
Under a deterministic design Under the current conditions $u_{\boldsymbol{a}_{i}} - u_{q_{i}} = O(1)$.
\end{lemma}
Proof. \cite{Barn:Cox:Infe:1994} state that if $(\hat{\eta}_{i},\boldsymbol{a}_{i})$ and $\boldsymbol{a}_{i}$ are approximately sufficient and ancillary to the order $O(n^{-1/2})$ then $p(\hat{\eta}_{i}|\eta_{i},\boldsymbol{a}_{i}) = p^{\dagger}(\hat{\eta}_{i}|\eta_{i},\boldsymbol{a}_{i}) + O(n^{-1})$, where $p^{\dagger}(\hat{\eta}_{i}|\eta_{i},\boldsymbol{a}_{i}) = \{\hat{\imath}_{\boldsymbol{y},\boldsymbol{t}_{i}}/(2\pi)\}^{1/2}e^{\overline{l}(\eta_{i},\hat{\eta}_{i})}$, where $\hat{\imath}_{\boldsymbol{y},\boldsymbol{t}_{i}}$ represents the observed information, $i_{\boldsymbol{y},\boldsymbol{t}_{i}}$, evaluated at $\eta_{i} = \hat{\eta}_{i}$. The limiting distribution of $q_{i}$ is normal with mean 0 and variance 1 which implies that it satisfies the $O(n^{-1/2})$ approximate sufficiency and ancillary condition. The importance of the $p^{\dagger}$-formula is, first, it implies that every $O(n^{-1/2})$ (or higher) approximate ancillary generates the same $p^{\dagger}$-formula and thus $u_{\boldsymbol{a}_{i}}^{\dagger} - u_{q_{i}}^{\dagger} = O(1)$. Condition \ref{cond:C2} ensures that the information in the relevant subset is accurately approximated by the $p^{\dagger}$-formula; specifically, $g_{\boldsymbol{a}_{i}} - g_{\boldsymbol{a}_{i}}^{\dagger} = O(1)$, where $g_{\boldsymbol{a}_{i}}^{\dagger} = {\rm{E}}^{\dagger}[i_{\boldsymbol{y},\boldsymbol{t}_{i}}]$. From its definition, in Theorem \ref{prop:LBexp}, the preceding implies that $u_{\boldsymbol{a}_{i}} -u_{\boldsymbol{a}_{i}}^{\dagger} = O(1)$ and by extension $u_{\boldsymbol{a}_{i}}  - u_{q_{i}} = O(1)$ as stated. $\square$

In words Lemma \ref{supp:lem:EHA} ensures that all approximate ancillaries yield the same approximate information. The implication begin that it only needs to be shown that ${\rm{E}}[\boldsymbol{u}_{\boldsymbol{Q}}\boldsymbol{v}_{\boldsymbol{Q}}^{T}] = \boldsymbol{0}$ and then an appeal to Lemma \ref{supp:lem:EHA} implies the stated result. This outlines the remainder of the proof. 

Condition \ref{cond:C2} ensures that $E[\dot{l}_{\boldsymbol{\mathcal{Y}},\boldsymbol{\mathcal{T}}_{\negmedspace i}}^{(\phi)} | \mathcal{Q}_{i}] = 0$; therefore,
\begin{align} \label{supp:eq:gQ}
     g_{\mathcal{Q}_{i}} =  {\rm{E}}[i_{\boldsymbol{\mathcal{Y}},\boldsymbol{\mathcal{T}}_{\negmedspace i}}^{(\phi)}|\mathcal{Q}_{i}] = \gamma({\rm{E}}[\mathcal{Q}_{i}|\mathcal{Q}_{i}] + nw_{i}) = \gamma(\mathcal{Q}_{i} + nw_{i}).
\end{align}
From its definition $\boldsymbol{u}_{\boldsymbol{Q}} = (u_{q_{1}},\ldots,u_{q_{d}})^{T}$, where $u_{q_{i}} = g_{q_{i}}  - w_{i}\sum_{i} g_{q_{i}}$ can be alternatively expressed as $u_{q_{i}} = \gamma\sum_{j=1}^{n}[t_{i}(j)-w_{i}][q_{y(j)}+1]$. Similarly, $\boldsymbol{v}_{\boldsymbol{Q}} = (v_{q_{1}},\ldots,v_{q_{d}})^{T}$, where $v_{q_{i}} = w_{i}(\sum_{i} g_{q_{i}} - n)$ can be alternatively expressed as $v_{q_{i}} = \gamma\sum_{j=1}^{n}q_{y(j)}$. 

Consider the variance and covariance terms of $u_{q_{i}}$
\begin{align}
    {\rm{Var}}[u_{\mathcal{Q}_{i}}] &= \gamma^{2}\sum_{j=1}^{n} \{t_{i}(j) - w_{i}\}^{2}{\rm{Var}}[q_{\mathcal{Y}(j)}] = \gamma^{2}\sum_{j=1}^{n} \{t_{i}(j) - w_{i}\}^{2} \\
    &= n\gamma^{2}w_{i}(1-w_{i}) \\
    {\rm{E}}[u_{\mathcal{Q}_{i}}u_{\mathcal{Q}_{k}}] &= \gamma^{2}\sum_{j=1}^{n}\sum_{l=1}^{n} \{t_{i}(j) - w_{i}\}\{t_{k}(l) - w_{k}\}{\rm{E}}[\{q_{\mathcal{Y}(j)}+1\}\{q_{\mathcal{Y}(l)}+1\}] \\
    &= \gamma^{2}\sum_{j=1}^{n}\{t_{i}(j) - w_{i}\}\{t_{k}(j) - w_{k}\} \\
    &= -n\gamma^{2}w_{i}w_{k}.
\end{align}
Note the independence property of $\boldsymbol{\mathcal{Y}}$ was used in the above. Next consider
\begin{align}
    {\rm{E}}[v_{\mathcal{Q}_{i}}v_{\mathcal{Q}_{k}}] &= \gamma^{2}w_{i}w_{k}{\rm{E}}\left[\left\{\sum_{j=1}^{n}q_{\mathcal{Y}(j)}\right\}^{2}\right] = w_{i}w_{k}\sum_{j=1}^{n}{\rm{Var}}[q_{\mathcal{Y}(j)}] = n\gamma^{2}w_{i}w_{k}
\end{align}
Now consider 
\begin{align}
    {\rm{E}}[u_{\mathcal{Q}_{i}}v_{\mathcal{Q}_{k}}] &= \gamma^{2}w_{k}\sum_{j=1}^{n}\sum_{l=1}^{n}\{t_{i}(j) - w_{i}\}{\rm{E}}\left[\{q_{\mathcal{Y}(j)}+1\}q_{\mathcal{Y}(l)}\right] \\
    &= \gamma^{2}w_{k} \sum_{j=1}^{n} \{t_{i}(j) - w_{i}\} =0,
\end{align}
where the independence of responses was again used in the above. The above implies that ${\rm{Var}}[\boldsymbol{u}_{\boldsymbol{\mathcal{Q}}}] = n\gamma^{2}\boldsymbol{W}(\boldsymbol{I}_{d} - \boldsymbol{w}\boldsymbol{w}^{T})$, ${\rm{Var}}[\boldsymbol{u}_{\boldsymbol{\mathcal{Q}}}] = n\gamma^{2}\boldsymbol{w}\boldsymbol{w}^{T}$ and ${\rm{E}}[\boldsymbol{u}_{\boldsymbol{Q}}\boldsymbol{v}_{\boldsymbol{Q}}^{T}] = \boldsymbol{0}$. This and Lemma \ref{supp:lem:EHA} implies that ${\rm{Var}}[\boldsymbol{u}_{\boldsymbol{\mathcal{A}}}] = n\gamma^{2}\boldsymbol{W}(\boldsymbol{I}_{d} - \boldsymbol{w}\boldsymbol{w}^{T})$, ${\rm{Var}}[\boldsymbol{u}_{\boldsymbol{\mathcal{A}}}] = n\gamma^{2}\boldsymbol{w}\boldsymbol{w}^{T}$ and ${\rm{E}}[\boldsymbol{u}_{\boldsymbol{A}}\boldsymbol{v}_{\boldsymbol{A}}^{T}] = \boldsymbol{0}$ hold at the limit as $n\rightarrow\infty$ for any approximate ancillary $\boldsymbol{\mathcal{A}}$.

\section{Proof of Lemma \ref{lem:Ameasurable}}
The following lemma is stated for random variables following the entire experiment. While this significantly more than is required these more general results imply the necessary results for the first two runs and will be useful for future results.
\begin{lemma} \label{supp:lem:Ameasurable}
$\boldsymbol{\mathcal{\check{A}}}\{1;s\}$ is $\boldsymbol{\mathcal{\check{A}}}\{1;r\}$ measurable for all $s \le r=1,\ldots,\mathcal{R}_{\max}$.
\end{lemma}

An important property of the location family is that $\boldsymbol{\mathcal{\check{A}}}\{1;s\}$ is $\boldsymbol{\mathcal{\check{A}}}\{1;r\}$ measurable for all $r\le s$ and $r=1,\dots,\mathcal{R}_{\max}$. To show this, the following notation is introduced. Note, from \citet{Fras:Anci:2004} the $\boldsymbol{\mathcal{\check{A}}}_{i}\{1;r\}$ can be written as 
\begin{align}
    \boldsymbol{\mathcal{\check{A}}}_{i}\{1;r\} &= \{\mathcal{\check{A}}_{i}\{1;r;1\},\ldots,\mathcal{\check{A}}_{i}\{1;r;n\{1\}\},\\
    &\quad\quad \ldots,\mathcal{\check{A}}_{i}\{1;r;\mathcal{N}\{1;r-1\}+1\},\ldots,\mathcal{\check{A}}_{i}\{1;r;\mathcal{N}_{i}\{1;r\}\}\}^{T},
\end{align}
where $\mathcal{\check{A}}_{i}\{1;r;j\}= y_{i}(j) - \overline{y}_{i}\{1;r\}$, $y_{i}(j)$ is the $j$th observation with support $x_{i}$ and $\overline{y}_{i}\{1;r\} = [\mathcal{N}\{1;r\}]^{-1}\sum_{i=1}^{n_{i}\{1;r\}}y_{i}(j)$ and $\boldsymbol{\check{A}}\{1;r\} = [\boldsymbol{\check{a}}_{1}\{1;r\},\ldots,\boldsymbol{\check{a}}_{d}\{1;r\}]$. To demonstrate this note $\mathcal{\check{A}}_{i}\{1;r;j\}$ can be written as
\begin{align}
    \mathcal{\check{A}}_{i}\{1;j\} = \mathcal{\check{A}}_{i}\{1;r;j\} - [n\{1\}]^{-1}\sum_{i=1}^{n\{1\}}\mathcal{\check{A}}_{i}\{1;r;j\}
\end{align}
for $j= 1,\ldots,n\{1\}$. Noting that the distribution of $n\{1\}$ is known  demonstrates that $\boldsymbol{a}_{i}\{1\}$ is $\boldsymbol{a}_{i}\{1;r\}$, measurable. Per the definition of the RRSD algorithm it can be seen that $\mathcal{N}_{i}\{2\}$ is $\boldsymbol{a}_{i}\{1\}$ measurable. Repeating as before we obtain
\begin{align}
    \mathcal{\check{A}}_{i}\{1,2;j\} = \mathcal{\check{A}}_{i}\{1;r;j\} - [\mathcal{N}_{i}\{1;2\}]^{-1}\sum_{i=1}^{\mathcal{N}_{i}\{1;2\}}\mathcal{\check{A}}_{i}\{1;r;j\}
\end{align}
for $j= 1,\ldots,\mathcal{N}_{i}\{1;2\}$; implying that $\boldsymbol{\check{a}}_{i}\{1;2\}$ is $\boldsymbol{\check{a}}_{i}\{1;r\}$, measurable. Repeating these steps sequentially, it can be shown that in general
\begin{align}
    \mathcal{\check{A}}_{i}\{1;s;j\} = \mathcal{\check{A}}_{i}\{1;r;j\} - [\mathcal{N}_{i}\{1;s\}]^{-1}\sum_{i=1}^{\mathcal{N}_{i}\{1;s\}}\mathcal{\check{A}}_{i}\{1;r;j\}
\end{align}
for $j= 1,\ldots,\mathcal{N}_{i}\{1;s\}$; from which it can be concluded that $\boldsymbol{\check{a}}_{i}\{1;s\}$ is $\boldsymbol{\check{a}}_{i}\{1;r\}$, measurable for all $s\le r$.

\section{Proof of Lemma \ref{lem:VarS}}

To begin recall the following lemma from \cite{Efro:Hink:Asse:1978}. 
\begin{lemma} \label{lem:efro:hink} Under the present conditions
\begin{align}
E[\hat{\eta}_{i} - \eta_{i}|\boldsymbol{\mathcal{A}}_{i}] &= -\hat{l}_{\boldsymbol{\mathcal{Y}},\boldsymbol{\mathcal{T}}_{\negmedspace i}}^{(\cdot 3)}/(2\hat{\imath}_{\boldsymbol{\mathcal{Y}},\boldsymbol{\mathcal{T}}_{\negmedspace i}}^{2})  + O_{p}(n^{-2}) \\
Var[\hat{\eta}_{i}|\boldsymbol{\mathcal{A}}_{i}] &= \hat{\imath}_{\boldsymbol{\mathcal{Y}},\boldsymbol{\mathcal{T}}_{\negmedspace i}}^{-1} + O_{p}(n^{-2}), \\
\hat{\imath}_{\boldsymbol{\mathcal{Y}},\boldsymbol{\mathcal{T}}_{\negmedspace i}}^{-1/2}(\hat{\eta}_{i} - \eta_{i})|\boldsymbol{\mathcal{A}}_{i} &\rightarrow N(0,1),
\end{align}
where convergence is in probability as $n\rightarrow\infty$.
\end{lemma}

The following lemma ensures that the conditional distribution of the MLE is the same as if no adaptation had taken place.
\begin{lemma}[\citet{lane2020efficiency} Theorem 5.1] \label{supp:lem:ancillary}
Let $(\boldsymbol{\check{\mathcal{Y}}},\boldsymbol{\mathcal{\check{T}}},\boldsymbol{\mathcal{\check{A}}})$ correspond to the responses, treatment allocation matrix and ancillary complement following an experiment using the RRSD; then
\begin{align} \label{eq:eqiv_dist}
f(\boldsymbol{\check{\mathcal{Y}}}|\check{\boldsymbol{\mathcal{T}}},\boldsymbol{\mathcal{\check{A}}}) \stackrel{d}{=} f(\boldsymbol{\mathcal{\check{Y}}}|\boldsymbol{\mathcal{\check{A}}})
\end{align}
where $f(\boldsymbol{\check{\mathcal{Y}}}|\boldsymbol{\check{A}})$ has the exact same form as an experiment with a deterministic design with $\boldsymbol{\mathcal{A}} = \boldsymbol{\mathcal{\check{A}}}$.
\end{lemma}

Proof. The result of the lemma is verified by the following sequence of a successive conditioning arguments. First,
\begin{align}
    f(\boldsymbol{\check{\mathcal{Y}}},\boldsymbol{\mathcal{\check{T}}},\boldsymbol{\mathcal{\check{A}}}) = f(\boldsymbol{\mathcal{\check{T}}}|\boldsymbol{\check{\mathcal{Y}}},\boldsymbol{\mathcal{\check{A}}}) f(\boldsymbol{\check{\mathcal{Y}}},\boldsymbol{\mathcal{\check{A}}}) = f(\boldsymbol{\mathcal{\check{T}}}|\boldsymbol{\mathcal{\check{A}}}) f(\boldsymbol{\check{\mathcal{Y}}},\boldsymbol{\mathcal{\check{A}}}).
\end{align}
The equality is due to the fact that the RRSD uses only information in $\boldsymbol{\mathcal{\check{A}}}$ to determine $\boldsymbol{\mathcal{\check{T}}}$.  Alternatively, we can also write
\begin{align}
    f(\boldsymbol{\check{\mathcal{Y}}},\boldsymbol{\mathcal{\check{T}}},\boldsymbol{\mathcal{\check{A}}}) &= f(\boldsymbol{\check{\mathcal{Y}}}|\boldsymbol{\mathcal{\check{T}}},\boldsymbol{\mathcal{\check{A}}})f(\boldsymbol{\mathcal{\check{T}}}|\boldsymbol{\mathcal{\check{A}}})f(\boldsymbol{\mathcal{\check{A}}}).
\end{align}
Combining the two results and re-arranging we obtain
\begin{align} \label{eq:DistEqual}
    f(\boldsymbol{\check{\mathcal{Y}}}|\boldsymbol{\mathcal{\check{T}}},\boldsymbol{\mathcal{\check{A}}}) = f(\boldsymbol{\check{\mathcal{Y}}}|\boldsymbol{\mathcal{\check{A}}}).
\end{align}
It can now be concluded that the treatment assignment implemented by the RRSD does not alter the conditional distribution of the responses. Of course the only way the RRSD influences the design is through the selection of the allocations; therefore we can conclude that the conditional distribution of the responses is the same as if $\boldsymbol{\mathcal{\check{T}}}$ were pre-specified at its observed value. $\square$

Lemmas \ref{lem:efro:hink} and \ref{supp:lem:ancillary} ensures the validity of the Taylor series of $i$ with respect to $\eta_{i}$ at $\hat{\eta}_{i}$
\begin{align} \label{eq:TS}
    i_{\boldsymbol{\mathcal{\check{Y}}}\{1;2\},\boldsymbol{\mathcal{\check{T}}}_{\negmedspace i}\{1;2\}} &= \hat{\imath}_{\boldsymbol{\mathcal{\check{Y}}}\{1;2\},\boldsymbol{\mathcal{\check{T}}}_{\negmedspace i}\{1;2\}} - \hat{l}_{\boldsymbol{\mathcal{\check{Y}}}\{1;2\},\boldsymbol{\mathcal{\check{T}}}_{\negmedspace i}\{1;2\}}^{(\cdot 3)}(\hat{\eta}_{i}\{1;2\} - \eta_{i}) \\
    &\quad- \hat{l}_{\boldsymbol{\mathcal{\check{Y}}}\{1;2\},\boldsymbol{\mathcal{\check{T}}}_{\negmedspace i}\{1;2\}}^{(\cdot 4)}(\hat{\eta}_{i}\{1;2\} - \eta_{i})^{2}/2 + o(1). 
\end{align}
Therefore, 
\begin{align}
    \mathcal{\check{S}}_{i}\{1\} = \mathcal{\hat{S}}_{i}\{1\} + \mathcal{\overline{B}}_{i}\{1;2\} - w_{i}\sum_{i=1}^{d}\mathcal{\overline{B}}_{i}\{1;2\}  + o(1),
\end{align}
where $\mathcal{\hat{S}}_{i}$ denotes $\mathcal{\check{S}}_{i}$ evaluated at the MLE of $\eta_{i}$ and $\mathcal{\overline{B}}_{i}\{1;2\} = (\hat{\eta}_{i}\{1;2\} - \eta_{i})\hat{l}_{\boldsymbol{\mathcal{\check{Y}}}\{1;2\},\boldsymbol{\mathcal{\check{T}}}_{\negmedspace i}\{1;2\}}^{(\cdot 3)} + (\hat{\eta}_{i}\{1;2\} - \eta_{i})^{2}\hat{l}_{\boldsymbol{\mathcal{\check{Y}}}\{1;2\},\boldsymbol{\mathcal{\check{T}}}_{\negmedspace i}\{1;2\}}^{(\cdot 4)}/2$. Noting that $\hat{l}_{\boldsymbol{\mathcal{\check{Y}}}\{1;2\},\boldsymbol{\mathcal{\check{T}}}_{\negmedspace i}\{1;2\}}^{(\cdot k)}$ is $\boldsymbol{\mathcal{\check{A}}}_{i}$ measurable for all $k=1,2,\ldots$ an application of Lemma \ref{lem:efro:hink} yields
\begin{align} \label{eq:Si1}
    {\rm{E}}[\mathcal{\check{S}}_{i}\{1\}|\boldsymbol{\mathcal{\check{A}}}_{i}\{1;2\}] = \mathcal{\hat{S}}_{i}\{1\} + \hat{B}_{i} - w_{i}\sum_{i=1}^{d}\hat{B}_{i} = \mathcal{\hat{S}}_{i}\{1\} + O(1),
\end{align}
where $\hat{B}_{i} = [(\hat{l}_{\boldsymbol{\mathcal{\check{Y}}}\{1;2\},\boldsymbol{\mathcal{\check{T}}}_{\negmedspace i}\{1;2\}}^{(\cdot 3)}/\hat{\imath}_{\boldsymbol{\mathcal{\check{Y}}}\{1;2\},\boldsymbol{\mathcal{\check{T}}}_{\negmedspace i}\{1;2\}})^{2} + \hat{l}_{\boldsymbol{\mathcal{\check{Y}}}\{1;2\},\boldsymbol{\mathcal{\check{T}}}_{\negmedspace i}\{1;2\}}^{(\cdot 4)}/\hat{\imath}_{\boldsymbol{\mathcal{\check{Y}}}\{1;2\},\boldsymbol{\mathcal{\check{T}}}_{\negmedspace i}\{1;2\}}]/2$. The $\boldsymbol{\mathcal{\check{A}}}_{i}\{1\}$-measurability of $\mathcal{S}_{i}\{1\}$ and \eqref{eq:Si1} implies that 
\begin{align}
    {\rm{Var}}[{\rm{E}}[\mathcal{S}_{i}\{1;2\}|\boldsymbol{\mathcal{\check{A}}}_{i}\{1;2\}]|\boldsymbol{\mathcal{\check{A}}}_{i}\{1\}] = {\rm{Var}}[{\rm{E}}[\mathcal{S}_{i}\{2\}|\boldsymbol{\mathcal{\check{A}}}_{i}\{1;2\}]|\boldsymbol{\mathcal{\check{A}}}_{i}\{1\}] + O(1)
\end{align}
as stated.

\section{Proof of Lemma \ref{lem:EH}}
All results are conditional on the initial design $\xi$. Further, all variables in this section pertain to the RRSD and thus the convention of using a check is not used since there should be no confusion.

Note that $(h_{\boldsymbol{\mathcal{\check{A}}}_{i}} - nw_{i}\mu)/\mu = \boldsymbol{u}_{\boldsymbol{\mathcal{\check{A}}}_{i}} + \boldsymbol{v}_{\boldsymbol{\mathcal{\check{A}}}_{i}}$. The following lemma implies Lemma \ref{lem:EH}. Note this lemma does not require the ancillary to be exact.

\begin{lemma} \label{supp:lem:EVEU} Under the present conditions ${\rm{E}}[\boldsymbol{v}_{\boldsymbol{\mathcal{\check{A}}}_{i}}] = 0$ and ${\rm{E}}[\boldsymbol{u}_{\boldsymbol{\mathcal{\check{A}}}}] = 0$.
\end{lemma}

For $v_{\boldsymbol{\mathcal{\check{A}}}_{i}}$ we find that
\begin{align}
    {\rm{E}}[v_{\boldsymbol{\mathcal{\check{A}}}_{i}}] &= w_{i}\sum_{i=1}^{d}({\rm{E}}[i_{\boldsymbol{\mathcal{\check{Y}}},\boldsymbol{\mathcal{\check{T}}}_{\negmedspace i}}^{(\phi)}] - n) = w_{i}\left({\rm{E}}\left[\sum_{i=1}^{d}\sum_{j=1}^{n}\mathcal{\check{T}}_{i}(j)i_{\mathcal{\check{Y}}(j)}^{(\phi)}\right] - n\right) \\
    &= w_{i}\sum_{j=1}^{n}({\rm{E}}\left[i_{\mathcal{\check{Y}}(j)}^{(\phi)}\right] - n) = 0, \label{eq:EVA}
\end{align}
for all $r=1,\ldots,\mathcal{R}_{\max}$. The first equality is from the definition of $v_{\boldsymbol{\mathcal{\check{A}}}_{i}}$; the second is from the definition of $i_{\boldsymbol{\check{y}},\boldsymbol{\check{t}}_{i}}$; the third is due to the fact that $\sum_{j=1}^{n}\mathcal{\check{T}}_{i}(j) = 1$; and the final equality is from ${\rm{E}}[i_{\mathcal{\check{Y}}(j)}^{(\phi)}] = 1$. 

For $u_{\boldsymbol{\mathcal{\check{A}}}_{i}}$, start by noting that 
\begin{align}
    {\rm{E}}\left[i_{\boldsymbol{\mathcal{\check{Y}}}\{r\},\boldsymbol{\mathcal{\check{T}}}_{\negmedspace i}\{r\} }^{(\phi)} \mid \boldsymbol{\mathcal{\check{A}}}\{1;r-1\}\right]  &={\rm{E}}\left[\left.\sum_{j=\mathcal{N}\{1;r-1\}+1}^{\mathcal{N}\{1;r\}}\mathcal{\check{T}}_{i}(j){\rm{E}}[i_{\mathcal{\check{Y}}(j)}^{(\phi)}|\mathcal{\check{T}}_{i}(j)]\right|\boldsymbol{\mathcal{\check{A}}}\{1;r-1\}\right] \\
    &= \sum_{j=\mathcal{N}\{1;r-1\}+1}^{\mathcal{N}\{1;r\}} {\rm{E}}\left[\mathcal{\check{T}}_{i}(j)|\boldsymbol{\mathcal{\check{A}}}\{1;r-1\}\right]\\
    &= \mathcal{N}\{r\} \mathcal{P}_{i}\{r\}. \label{eq:AirEqual}
\end{align}
The first equality is from the definition of $i_{\boldsymbol{\mathcal{\check{Y}}}\{r\},\boldsymbol{\mathcal{\check{T}}}_{\negmedspace i}\{r\} }^{(\phi)}\{r\}$; the second is due to the fact that, conditional on $\mathcal{\check{T}}_{i}(j)$, $i_{\mathcal{\check{Y}}(j)}^{(\phi)}$ is independent of the past with unit expectation and that $\mathcal{N}\{1;r\}$ is $\boldsymbol{\mathcal{\check{A}}}\{1;r-1\}$ measurable; the final equality is due to the definition of $\mathcal{P}_{i}\{r\} = {\rm{E}}[\mathcal{\check{T}}_{i}(j)|\boldsymbol{\mathcal{\check{A}}}\{1;r-1\}]$ for all $j=\mathcal{N}\{1;r-1\}+1,\ldots,\mathcal{N}\{1;r\}$. 
Consider the conditional expectation
\begin{align}
   &{\rm{E}}[u_{\boldsymbol{\mathcal{\check{A}}}_{i}\negmedspace\{1;r\}}| \boldsymbol{\mathcal{\check{A}}}\{1;r-1\}] \\
   &= {\rm{E}}\left[{\rm{E}}[\mathcal{\check{S}}_{i}\{1;r\} \mid \boldsymbol{\mathcal{\check{A}}}_{i}\negmedspace\{1;r\} ] \mid \boldsymbol{\mathcal{\check{A}}}\{1;r-1\} \right] \\
   &= u_{\boldsymbol{\mathcal{\check{A}}}_{i}\negmedspace\{1;r-1\}} + {\rm{E}} \left[\mathcal{\check{S}}_{i}\{r\} \mid \boldsymbol{\mathcal{\check{A}}}\{1;r-1\}] \right] \\
   &= u_{\boldsymbol{\mathcal{\check{A}}}_{i}\negmedspace\{1;r-1\}} +  {\rm{E}}\left[\left.i_{\boldsymbol{\mathcal{\check{Y}}}\{r\},\boldsymbol{\mathcal{\check{T}}}_{\negmedspace i}\{r\} }^{(\phi)} - w_{i} \sum_{i=1}^{d}i_{\boldsymbol{\mathcal{\check{Y}}}\{r\},\boldsymbol{\mathcal{\check{T}}}_{\negmedspace i}\{r\} }^{(\phi)} \right | \boldsymbol{\mathcal{\check{A}}}\{1;r-1\} \right] \\
   &= u_{\boldsymbol{\mathcal{\check{A}}}_{i}\negmedspace\{1;r-1\}} + \mathcal{N}\{r\} \left\{ \mathcal{P}_{i}\{r\} - w_{i} \sum_{j=1}^{d}  \mathcal{P}_{i}\{r\} \right\}  \\
   &= u_{\boldsymbol{\mathcal{\check{A}}}_{i}\negmedspace\{1;r-1\}} + \mathcal{N}\{r\} \{\mathcal{P}_{i}\{r\} - w_{i}\} . \label{eq:EMuair}
\end{align}
The first equality is from the definition of $u_{\boldsymbol{\mathcal{\check{A}}}_{i}\negmedspace\{1;r\}}$; the second follows from the fact that $u_{\boldsymbol{\mathcal{\check{A}}}_{i}\negmedspace\{1;r-1\}}$ and $\boldsymbol{\mathcal{\check{A}}}_{i}\negmedspace\{1;r-1\}$ are $\boldsymbol{\mathcal{\check{A}}}_{i}\negmedspace\{1;r\}$ measurable; the third is from the definition $\mathcal{\check{S}}_{i}\{r\}$; the fourth is from \eqref{eq:AirEqual}; the final line is obtained by recognizing that $\sum_{j=1}^{d}  \mathcal{P}_{i}\{r\} = 1$. 

Let $\mathcal{G}\{r\} = I[\mathcal{M}\{r\}>n-\mathcal{N}\{1;r-1\}]$. From Algorithm \ref{alg:RRSD} the definition of $\mathcal{P}_{i}\{r\}$ differs based on whether $\mathcal{G}\{r\}$ is 0 or 1. Note that $\mathcal{G}\{r\}$ is $\boldsymbol{\mathcal{\check{A}}}\{1;r-1\}$ measurable. First, for the case where $\mathcal{G}\{r\}=0$ we obtain
\begin{align}
    {\rm{E}}[\mathcal{\check{S}}_{i}\{1;r\}| \mathcal{G}\{r\}=0]
    = {\rm{E}}[u_{\boldsymbol{\mathcal{\check{A}}}_{i}\negmedspace\{1;r\}}| \boldsymbol{\mathcal{\check{A}}}\{1;r-1\},\mathcal{G}\{r\}=0] =  0. \label{eq:ExpUi0}
\end{align}
This is obtained directly by recognizing that if $\mathcal{G}\{r\}=0$ then $\mathcal{P}_{i}\{r\} = w_{i} - u_{\boldsymbol{\mathcal{\check{A}}}_{i}\negmedspace\{1;r-1\}}/ \mathcal{N}\{r\}$ and plugging this fact in to \eqref{eq:EMuair}. Second, for the case where $\mathcal{G}\{r\}=1$ we obtain
\begin{align}
    &{\rm{E}}[u_{\boldsymbol{\mathcal{\check{A}}}_{i}\negmedspace\{1;r\}}| \boldsymbol{\mathcal{\check{A}}}\{1;r-2\},\mathcal{G}\{r\}=1] \\
    &= {\rm{E}}[ {\rm{E}}[\mathcal{\check{S}}_{i}\{1;r\}|\boldsymbol{\mathcal{\check{A}}}_{i}\negmedspace\{1;r\},\mathcal{G}\{r\}=1] | \boldsymbol{\mathcal{\check{A}}}\{1;r-2\},\mathcal{G}\{r\}=1] \\
    &= {\rm{E}}[\mathcal{\check{S}}_{i}\{1;r-1\} | \boldsymbol{\mathcal{\check{A}}}\{1;r-2\},\mathcal{G}\{r\}=1] \\
    &= {\rm{E}}[u_{\boldsymbol{\mathcal{\check{A}}}_{i}\negmedspace\{1;r-1\}}| \boldsymbol{\mathcal{\check{A}}}\{1;r-2\},\mathcal{G}\{r-1\}=0] \\
    &= 0. \label{eq:ExpUi1}
\end{align}
The first line is due to successive conditioning; the second is obtained by recognizing that if $\mathcal{G}\{r\}=1$ then $\mathcal{P}_{i}\{r\} = w_{i}$ and ${\rm{E}}[\mathcal{\check{S}}_{i}\{r\} \mid \mathcal{G}\{r\}=1] = {\rm{E}}[\mathcal{\check{S}}_{i}\{r\} \mid \mathcal{G}\{r\}=1]=0$; the third line is again from successive conditioning, noting that if $\mathcal{G}\{r\}=1$ then $\mathcal{G}\{r-1\}=1$ and the definition of $u_{\boldsymbol{\mathcal{\check{A}}}_{i}\negmedspace\{1;r-1\}}$; the final equality is from \eqref{eq:ExpUi0}. Equations \eqref{eq:ExpUi0} and \eqref{eq:ExpUi1} imply that
\begin{align}
    {\rm{E}}[u_{\boldsymbol{\mathcal{\check{A}}}_{i}\negmedspace\{1;r\}}|\mathcal{G}\{r\}=0] &= {\rm{E}}[\mathcal{\check{S}}_{i}\{1;r\}|\mathcal{G}\{r\}=0] = 0 \quad\mbox{and} \label{eq:Ui0}\\ 
   {\rm{E}}[u_{\boldsymbol{\mathcal{\check{A}}}_{i}\negmedspace\{1;r\}}|\mathcal{G}\{r\}=1] &= {\rm{E}}[\mathcal{\check{S}}_{i}\{1;r\}|\mathcal{G}\{r\}=1] = 0. \label{eq:Ui1}
\end{align}
The above verifies that
\begin{align}
     {\rm{E}}[u_{\boldsymbol{\mathcal{\check{A}}}_{i}\negmedspace\{1;r\}}] &= 0
\end{align}
for all $r\ge2$. Note, it is very straightforward to verify that ${\rm{E}}[u_{\boldsymbol{\mathcal{\check{A}}}_{i}\negmedspace\{1\}}] = 0$. This verifies that $ {\rm{E}}[u_{\boldsymbol{\mathcal{\check{A}}}_{i}\negmedspace\{1;r\}}] = 0$ for all $r$ and implies the stated result. 

\section{Proof of Lemma \ref{lem:VV}} 

To begin note that Lemma implies that the following holds for fixed designs and Lemma \ref{supp:lem:ancillary} ensures that it extends to the RRSD
\begin{align}
    {\rm{E}}[i_{\boldsymbol{\mathcal{\check{Y}}},\boldsymbol{\mathcal{\check{T}}}_{\negmedspace i}}^{(\phi)} - n|\boldsymbol{\mathcal{\check{A}}}] &= {\rm{E}}[i_{\boldsymbol{\mathcal{\check{Y}}},\boldsymbol{\mathcal{\check{T}}}_{\negmedspace i}}^{(\phi)} - n + \nu_{11}\dot{l}_{\boldsymbol{\mathcal{\check{Y}}},\boldsymbol{\mathcal{\check{T}}}_{\negmedspace i}}^{(\phi)}|\boldsymbol{\mathcal{\check{A}}}] = {\rm{E}}[\mathcal{\check{Q}}_{i}|\boldsymbol{\mathcal{\check{A}}}] \\
    &= {\rm{E}}[\mathcal{\check{Q}}_{i}|\boldsymbol{\mathcal{\check{Q}}}] + O(1) = \mathcal{\check{Q}}_{i} + O(1).
\end{align}
Next recall that  $l_{\eta}^{(\cdot k)}(y) = [\partial^{k}/\partial \eta^{k}] \log f_{\eta}(y)$ is the $k$th derivative log likelihood for a single observation. The location condition implies that $l_{\eta}^{(\cdot k)}(y) = l_{0}^{(\cdot k)}(\varepsilon)$, where $l_{0}^{(\cdot k)}(\varepsilon) = (-1)^{k}[\partial^{k}/\partial \varepsilon^{k}] \log f_{0}(\varepsilon)$. The importance of this is to say that log-likelihood and its derivatives depend only on the errors and not on the responses. The error distribution is unaffected by the design. The consequence of this is that since $\sum_{i=1}^{d}\mathcal{\check{Q}}_{i} = \sum_{j=1}^{n}q_{\mathcal{\check{Y}}(j)}$ is a function of the first and second derivative of the log-likelihood is actually only a function of only the errors and thus its distribution is exactly the same as if no adaptation had taken place. This implies that
\begin{align}
    n^{-1}\lim_{n\rightarrow\infty} {\rm{Var}}[v_{\boldsymbol{\mathcal{\check{A}}}_{i}}] = n^{-1}\lim_{n\rightarrow\infty} {\rm{Var}}[\sum_{i=1}^{d}\mathcal{\check{Q}}_{i}] = \gamma^{2}
\end{align}
as stated. 
\section{Proof of Lemma \ref{lem:VU} }

To begin we write 
\begin{align}
    {\rm{Var}}[u_{\boldsymbol{\mathcal{\check{A}}}_{i}\{1;r\}}] &= {\rm{Var}}[{\rm{E}}[u_{\boldsymbol{\mathcal{\check{A}}}_{i}\{1;r\}}|\mathcal{G}\{r\}]] + {\rm{E}}[{\rm{Var}}[u_{\boldsymbol{\mathcal{\check{A}}}_{i}\{1;r\}}|\mathcal{G}\{r\}]] \\
    &= {\rm{E}}[{\rm{Var}}[u_{\boldsymbol{\mathcal{\check{A}}}_{i}\{1;r\}}|\mathcal{G}\{r\}]] \\
    &= {\rm{Var}}[u_{\boldsymbol{\mathcal{\check{A}}}_{i}\{1;r\}}|\mathcal{G}\{r\}=0]P\{\mathcal{G}\{r\}=0\} \\
    &\quad + {\rm{Var}}[u_{\boldsymbol{\mathcal{\check{A}}}_{i}\{1;r\}}|\mathcal{G}\{r\}=1]P\{\mathcal{G}\{r\}=1\}. \label{eq:VUAFull}
\end{align}
The first line is from the law of total variance; the second follows directly from \eqref{eq:Ui0} and \eqref{eq:Ui1}; and the third is from the definition of variance. Consider the term
\begin{align}
    {\rm{Var}}[u_{\boldsymbol{\mathcal{\check{A}}}_{i}\{1;r\}}|\mathcal{G}\{r\}=0] &= {\rm{Var}}[{\rm{E}}[u_{\boldsymbol{\mathcal{\check{A}}}_{i}\{1;r\}}|\boldsymbol{\mathcal{\check{A}}}_{i}\{1;r-1\},\mathcal{G}\{r\}=0] | \mathcal{G}\{r\}=0] \\
    &\quad= {\rm{E}}[{\rm{Var}}[u_{\boldsymbol{\mathcal{\check{A}}}_{i}\{1;r\}}|\boldsymbol{\mathcal{\check{A}}}_{i}\{1;r-1\},\mathcal{G}\{r\}=0] | \mathcal{G}\{r\}=0] \\
    &= {\rm{E}}[{\rm{Var}}[u_{\boldsymbol{\mathcal{\check{A}}}_{i}\{1;r\}}|\boldsymbol{\mathcal{\check{A}}}_{i}\{1;r-1\},\mathcal{G}\{r\}=0] | \mathcal{G}\{r\}=0]. \label{eq:VUG0}
\end{align}
The first line is from the law of total variance and the second follows directly from \eqref{eq:Ui0}. Similarly, the term 
\begin{align}
    {\rm{Var}}[u_{\boldsymbol{\mathcal{\check{A}}}_{i}\{1;r\}}|\mathcal{G}\{r\}=1] &= {\rm{Var}}[{\rm{E}}[u_{\boldsymbol{\mathcal{\check{A}}}_{i}\{1;r\}}|\boldsymbol{\mathcal{\check{A}}}_{i}\{1;r-2\},\mathcal{G}\{r\}=1] | \mathcal{G}\{r\}=1] \\
    &\quad= {\rm{E}}[{\rm{Var}}[u_{\boldsymbol{\mathcal{\check{A}}}_{i}\{1;r\}}|\boldsymbol{\mathcal{\check{A}}}_{i}\{1;r-2\},\mathcal{G}\{r\}=1] | \mathcal{G}\{r\}=1] \\
    &= {\rm{E}}[{\rm{Var}}[u_{\boldsymbol{\mathcal{\check{A}}}_{i}\{1;r\}}|\boldsymbol{\mathcal{\check{A}}}_{i}\{1;r-2\},\mathcal{G}\{r\}=1] | \mathcal{G}\{r\}=1]. \label{eq:VUG1}
\end{align}
The first line is from the law of total variance and the second follows directly from \eqref{eq:Ui1}. The following lemma provides a bound on both \eqref{eq:VUG0} and \eqref{eq:VUG1}. 
\begin{lemma} \label{lem:VUA}
Under the current conditions 
\begin{align}
    &{\rm{Var}}[u_{\boldsymbol{\mathcal{\check{A}}}_{i}\{1;r\}}|\boldsymbol{\mathcal{\check{A}}}_{i}\{1;s\},\mathcal{G}\{r\} = g\{r\}] \\
    &\quad= {\rm{Var}}[E[ \mathcal{\check{S}}_{i}\{s+1\}+\ldots+\mathcal{\check{S}}_{i}\{r\}| \boldsymbol{\mathcal{\check{A}}}_{i}\{1;r\} ]|\boldsymbol{\mathcal{\check{A}}}_{i}\{1;s\},\mathcal{G}\{r\} = g\{r\}] + O_{p}(1)
\end{align}
for $s=r-1$ and $r-2$, where $g\{r\} = 0$ or 1.
\end{lemma}
The steps of the proof of Lemma \ref{lem:VarS} can effectively be repeated with a simple change of notation to incorporate the additional runs. Due to the similarity with what has already been shown the proof of Lemma \ref{lem:VUA} has been omitted. 

Additionally, from an application of the law of total variance
\begin{align}
    &{\rm{Var}}[E[ \mathcal{\check{S}}_{i}\{s+1\}+\ldots+\mathcal{\check{S}}_{i}\{r\}| \boldsymbol{\mathcal{\check{A}}}_{i}\{1;r\} ]|\boldsymbol{\mathcal{\check{A}}}_{i}\{1;s\},\mathcal{G}\{r\} = g\{r\}] \\
    &\quad\le {\rm{Var}}[\mathcal{\check{S}}_{i}\{s+1\}+\ldots+\mathcal{\check{S}}_{i}\{r\}|\boldsymbol{\mathcal{\check{A}}}_{i}\{1;s\},\mathcal{G}\{r\} = g\{r\}].
\end{align}
Therefore \eqref{eq:VUAFull} can be bounded as 
\begin{align}
    \lim_{n\rightarrow\infty} n^{-1/2}{\rm{Var}}[u_{\boldsymbol{\mathcal{\check{A}}}_{i}\{1;r\}}] &\le  \lim_{n\rightarrow\infty} n^{-1}{\rm{E}}[(1 - \mathcal{G}\{r\}){\rm{Var}}[\mathcal{\check{S}}_{i}\{r\}|\boldsymbol{\mathcal{\check{A}}}_{i}\{1;r-1\}]] \\
    &\quad+ \lim_{n\rightarrow\infty} n^{-1/2}{\rm{E}}[\mathcal{G}\{r\}{\rm{Var}}[\mathcal{\check{S}}_{i}\{r\} + \mathcal{\check{S}}_{i}\{r-1\}|\boldsymbol{\mathcal{\check{A}}}_{i}\{1;r-2\}]] \\
    &= \lim_{n\rightarrow\infty} n^{-1/2}{\rm{E}}[(1 - \mathcal{G}\{r\}){\rm{Var}}[\mathcal{\check{S}}_{i}\{r\}|\boldsymbol{\mathcal{\check{A}}}_{i}\{1;r-1\}]] \label{limVUA1}\\
    &\quad+ \lim_{n\rightarrow\infty} n^{-1/2}{\rm{E}}[\mathcal{G}\{r\}{\rm{Var}}[\mathcal{\check{S}}_{i}\{r\}|\boldsymbol{\mathcal{\check{A}}}_{i}\{1;r-2\}]] \label{limVUA2}\\
    &\quad+ \lim_{n\rightarrow\infty} n^{-1/2}{\rm{E}}[\mathcal{G}\{r\}{\rm{Var}}[ \mathcal{\check{S}}_{i}\{r-1\}|\boldsymbol{\mathcal{\check{A}}}_{i}\{1;r-2\}]] \label{limVUA3}\\
    &\quad+ \lim_{n\rightarrow\infty} n^{-1/2}{\rm{E}}[\mathcal{G}\{r\}{\rm{Cov}}[\mathcal{\check{S}}_{i}\{r\}, \mathcal{\check{S}}_{i}\{r-1\}|\boldsymbol{\mathcal{\check{A}}}_{i}\{1;r-2\}]] \label{limVUA4}.
\end{align}
Consider the interior of the expectation in \eqref{limVUA2}
\begin{align}
    &{\rm{Var}}[\mathcal{\check{S}}_{i}\{r\}|\boldsymbol{\mathcal{\check{A}}}\{1;r-2\},\mathcal{G}\{r\}=1] \\
    &\quad\quad= {\rm{Var}}\left[\sum_{j=\mathcal{N}\{1;r-1\}+1}^{\mathcal{N}\{1;r\}}i_{\mathcal{\check{Y}}(j)}^{(\phi)}[\mathcal{\check{T}}_{i}(j) - w_{i}]\mid \boldsymbol{\mathcal{Q}}\{1;r-2\},\mathcal{G}\{r\}=1\right] \\
    &\quad\quad=\sum_{j=\mathcal{N}\{1;r-1\}+1}^{\mathcal{N}\{1;r\}}\left({\rm{E}}[\{\mathcal{\check{T}}_{i}(j) - w_{i}\}{\rm{Var}}\left[i_{\mathcal{\check{Y}}(j)}^{(\phi)}\mid\mathcal{\check{T}}_{i}(j),\mathcal{G}\{r\}=1\right]\mid \mathcal{G}\{r\}=1 ] \right.\\
    &\quad\quad= \sum_{j=\mathcal{N}\{1;r-1\}+1}^{\mathcal{N}\{1;r\}}{\rm{E}}[\{\mathcal{\check{T}}_{i}(j) - w_{i}\}^{2}\mid \mathcal{G}\{r\}=1 ] \\
    &\quad\quad= a_{1}\mathcal{N}\{r\}. \label{eq:condVS}
\end{align}
where $a_{1} = w_{i}(1-w_{i})$. One important aspect of the preceding derivations is that $\mathcal{G}\{r\}=1$ implies that $r$ is the final run and that $\mathcal{P}_{i}\{r\} = w_{i}$ are fixed. Additionally, this implies that given $\mathcal{G}\{r\}=1$ that $\mathcal{\check{T}}_{i}(j)$ are independent Bernoulli random variables with probability $w_{i}$. The first equality is obtained from the definition of $\mathcal{\check{S}}_{i}\{r\}$; the second equality is fomr the preceding comment that given $\mathcal{G}\{r\}=1$ then $r$ is the final run and that $\mathcal{N}\{r\} = n-\mathcal{N}\{1;r-1\}$ is $\boldsymbol{\mathcal{\check{A}}}\{1;r-2\}$ measurable and that $i_{\mathcal{\check{Y}}(j)}^{(\phi)}$ is a sum $\mathcal{N}\{r\}$ independent random variables; the third equality is obtained by recalling that ${\rm{E}}[i_{\mathcal{\check{Y}}(j)}^{(\phi)}] =1$, ${\rm{Var}}[i_{\mathcal{\check{Y}}(j)}^{(\phi)}] = \gamma^{2}$ and as previously described, given $\mathcal{G}\{r\}=1$ that $\mathcal{\check{T}}_{i}(j)$ are independent Bernoulli random variables with probability $w_{i}$; the final equality is by simplifying terms. 

Using \eqref{eq:condVS} equation \eqref{limVUA2} can be written as
\begin{align}
    {\rm{E}}[{\rm{Var}}[\mathcal{\check{S}}_{i}\{r\}|\boldsymbol{\mathcal{\check{A}}}\{1;r-2\},\mathcal{G}\{r\}=1]|\mathcal{G}\{r\}=1]P\{\mathcal{G}\{r\}=1\} &= a_{1}{\rm{E}}[\mathcal{G}\{r\}\mathcal{N}\{r\}].
\end{align}

Consider the interior of the expectation in \eqref{limVUA1} and recall that $i_{\mathcal{\check{Y}}(j)}^{(\phi)}[\mathcal{\check{T}}_{i}(j) - w_{i}]$, $j=\mathcal{N}\{1;r-1\},\ldots,\mathcal{N}\{1;r\}$ given $\boldsymbol{\mathcal{\check{A}}}_{i}\negmedspace\{1;r-1\}$ is a sequence of $\mathcal{N}\{r\}$ independent and identically distributed random variables; therefore,
\begin{align}
    {\rm{Var}}[\mathcal{\check{S}}_{i}\{r\}|\boldsymbol{\mathcal{\check{A}}}_{i}\negmedspace\{1;r-1\}] 
    &= \sum_{j=\mathcal{N}\{1;r-1\}+1}^{\mathcal{N}\{1;r\}} {\rm{Var}}\left[  i_{\mathcal{\check{Y}}(j)}^{(\phi)}[\mathcal{\check{T}}_{i}(j) - w_{i}] |\boldsymbol{\mathcal{\check{A}}}_{i}\negmedspace\{1;r-1\}\right],
\end{align}
where
\begin{align}
    &{\rm{Var}}\left[i_{\mathcal{\check{Y}}(j)}^{(\phi)}[\mathcal{\check{T}}_{i}(j) - w_{i}]|\boldsymbol{\mathcal{\check{A}}}_{i}\negmedspace\{1;r-1\}\right] \\
    &\quad\quad= {\rm{E}}\left[{\rm{Var}}\left[i_{\mathcal{\check{Y}}(j)}^{(\phi)}[\mathcal{\check{T}}_{i}(j) - w_{i}]|\mathcal{\check{T}}_{i}(j)\right]|\boldsymbol{\mathcal{\check{A}}}_{i}\negmedspace\{1;r\}\right] \\
    &\quad\quad\quad\quad + {\rm{Var}}\left[{\rm{E}}\left[i_{\mathcal{\check{Y}}(j)}^{(\phi)}[\mathcal{\check{T}}_{i}(j) - w_{i}]|\mathcal{\check{T}}_{i}(j)\right]|\boldsymbol{\mathcal{\check{A}}}_{i}\negmedspace\{1;r-1\}\right]  \\
    &\quad\quad= \gamma^{2}{\rm{E}}[(\mathcal{\check{T}}_{i}(j) - w_{i})^{2}|\boldsymbol{\mathcal{\check{A}}}_{i}\negmedspace\{1;r-1\}] + {\rm{Var}}[\mathcal{\check{T}}_{i}(j)|\boldsymbol{\mathcal{\check{A}}}_{i}\negmedspace\{1;r-1\}] \\
    &\quad\quad= \gamma^{2}{\rm{E}}[\mathcal{\check{T}}_{i}(j)^{2} - 2\mathcal{\check{T}}_{i}(j)w_{i} + w_{i}^{2}|\boldsymbol{\mathcal{\check{A}}}_{i}\negmedspace\{1;r-1\}] + \mathcal{P}_{i}\{r\}[1-\mathcal{P}_{i}\{r\}] \\
    &\quad\quad= \gamma^{2}\{{\rm{Var}}[\mathcal{\check{T}}_{i}(j)|\boldsymbol{\mathcal{\check{A}}}_{i}\negmedspace\{1;r-1\}] + {\rm{E}}[\mathcal{\check{T}}_{i}(j)|\boldsymbol{\mathcal{\check{A}}}_{i}\negmedspace\{1;r-1\}]^{2}\\
     &\quad\quad\quad\quad- 2\mathcal{P}_{i}\{r\}w_{i} + w_{i}^{2}\} + \mathcal{P}_{i}\{r\}[1-\mathcal{P}_{i}\{r\}]\\
    &\quad\quad= \gamma^{2}\{\mathcal{P}_{i}\{r\}[1-\mathcal{P}_{i}\{r\}]+ \mathcal{P}_{i}^{2}\{r\} - 2\mathcal{P}_{i}\{r\}w_{i} + w_{i}^{2}\} + \mathcal{P}_{i}\{r\}[1-\mathcal{P}_{i}\{r\}] \\
    &\quad\quad\le a_{2},
\end{align}
where $a_{2} =  (\gamma^{2} + 1)/4$. Using the above we obtain
\begin{align}
    {\rm{Var}}[\mathcal{\check{S}}_{i}\{r\}|\boldsymbol{\mathcal{\check{A}}}_{i}\negmedspace\{1;r-1\}] \le a_{2}\mathcal{N}\{r\}.
\end{align}
The above was derived for an arbitrary $r$; therefore, \eqref{limVUA3} can be similarly bounded as
\begin{align}
    {\rm{Var}}[\mathcal{\check{S}}_{i}\{r-1\}|\boldsymbol{\mathcal{\check{A}}}_{i}\negmedspace\{1;r-2\}] \le a_{2}\mathcal{N}\{r-1\}.
\end{align}
For the covariance term in \eqref{limVUA4} first recognize that
\begin{align}
    &{\rm{E}}[\mathcal{\check{S}}_{i}\{r\}\mathcal{\check{S}}_{i}\{r-1\}|\boldsymbol{\mathcal{\check{A}}}\{1;r-2\},\mathcal{G}\{r\}=1] \\
    &\quad\quad= {\rm{E}}[\mathcal{\check{S}}_{i}\{r-1\}E[\mathcal{\check{S}}_{i}\{r\}|\boldsymbol{\mathcal{\check{A}}}_{i}\negmedspace\{1;r-1\},\mathcal{G}\{r\}=1]|\boldsymbol{\mathcal{\check{A}}}\{1;r-2\},\mathcal{G}\{r\}=1] \\ &\quad\quad= 0.
\end{align}
The first line is from successive conditioning and the second is due to the fact that ${\rm{E}}[\mathcal{\check{S}}_{i}\{r\}|\boldsymbol{\mathcal{\check{A}}}_{i}\negmedspace\{1;r-1\},\mathcal{G}\{r\}=1]=0$. The preceding implies that 
\begin{align}
    {\rm{Cov}}[\mathcal{\check{S}}_{i}\{r\},\mathcal{\check{S}}_{i}\{r-1\}|\boldsymbol{\mathcal{\check{A}}}\{1;r-2\},\mathcal{G}\{r\}=1]=0
\end{align}
which in turn implies that \eqref{limVUA4} equals 0.

Finally, \eqref{eq:VUAFull} can be bounded as
\begin{align} 
    &\lim_{n\rightarrow\infty} n^{-1/2}{\rm{Var}}[u_{\boldsymbol{\mathcal{\check{A}}}_{i}\{1;r\}}]\\
    &\le \lim_{n\rightarrow\infty} n^{-1/2}\left[  a_{2}{\rm{E}}[[1-\mathcal{G}\{r\}]\mathcal{N}\{r\}]
    + a_{1}{\rm{E}}[\mathcal{G}\{r\}\mathcal{N}\{r\}] + a_{2}{\rm{E}}[\mathcal{G}\{r\}\mathcal{N}\{r-1\}] \right]\\
    &\le \lim_{n\rightarrow\infty} n^{-1/2} \left[(a_{1}+a_{2}){\rm{E}}[\mathcal{N}\{r\}] + a_{2}{\rm{E}}[\mathcal{N}\{r-1\}] \right], \label{eq:VUABound}
\end{align}
since $a_{1},a_{2}>0$. 

A useful expression for bounding the right hand side of the above is
\begin{align}
     {\rm{E}}[\mathcal{N}\{r\}] &\le {\rm{E}}\left[\max_{i}\left\{u_{\boldsymbol{\mathcal{\check{A}}}_{i}\{1;r-1\}}/w_{i}\right\} + 1\right] \\
     &\le \left({\rm{E}}\left[\left|\max_{i} \left\{u_{\boldsymbol{\mathcal{\check{A}}}_{i}\{1;r-1\}}/w_{i}\right\}\right|^{2}\right]\right)^{1/2}  + 1 \\
     &\le \left({\rm{E}}\left[\max_{i}\left\{(u_{\boldsymbol{\mathcal{\check{A}}}_{i}\{1;r-1\}}/w_{i})^{2}\right\}\right]\right)^{1/2}  + 1 \\
     &\le \left({\rm{E}}\left[\sum_{i=1}^{d}\{u_{\boldsymbol{\mathcal{\check{A}}}_{i}\{1;r-1\}}/w_{i}\}^{2}\right]\right)^{1/2}  + 1 \\
     &\le \left(\sum_{i=1}^{d}w_{i}^{-2}{\rm{E}}\left[u_{\boldsymbol{\mathcal{\check{A}}}_{i}\{1;r-1\}}^{2}\{1;r-1\}\right]\right)^{1/2}  + 1 \\
     &= w_{\min}^{-1} \left(\sum_{i=1}^{d}{\rm{Var}}\left[u_{\boldsymbol{\mathcal{\check{A}}}_{i}\{1;r-1\}}\right]\right)^{1/2}  + 1 
\end{align}
which holds for all $r=3,\ldots,\mathcal{R}_{\max}$, where $w_{\min} = \min_{i}w_{i} > 0$ . The first line is from the definition of $\mathcal{N}\{r\}$ in Algorithm \ref{alg:RRSD}, where the $+1$ is to account for the ceiling operator; the second line is due to Holder; the third is from a property of maximum of a list; the fourth is from the fact that the maximum of a list of squares is less than the total of the squares; the fifth is obtained by rearranging terms; the sixth is a result of the fact that $E\left[u_{\boldsymbol{\mathcal{\check{A}}}_{i}}\right] = 0$.

Consider that
\begin{align}
     &\lim_{n\rightarrow\infty} n^{-1/2} \sum_{i=1}^{d} {\rm{Var}}[u_{\boldsymbol{\mathcal{\check{A}}}_{i}\negmedspace\{1;r\}}] \\
     &\quad\quad\le \lim_{n\rightarrow\infty} n^{-1/2} \left[(a_{1}+a_{2})w_{\min}^{-1} \left(\sum_{i=1}^{d}{\rm{Var}}\left[u_{\boldsymbol{\mathcal{\check{A}}}_{i}\{1;r-1\}}\right]\right)^{1/2} \right. \\
     &\quad\quad\quad\quad\left.+ a_{2}w_{\min}^{-1} \left(\sum_{i=1}^{d}{\rm{Var}}\left[u_{\boldsymbol{\mathcal{\check{A}}}_{i}\{1;r-2\}}\right]\right)^{1/2}\right] \\
     &\quad\quad\le \lim_{n\rightarrow\infty} n^{-1/2}d(a_{1}+a_{2})\left(w_{\min}^{-1}\left\{d(a_{1} + a_{2})E[\mathcal{N}\{r-1\}]
    + da_{2}E[\mathcal{N}\{r-2\}]\right\}^{1/2} \right) \\
    &\quad\quad\quad\quad+ n^{-1/2}da_{2}\left(w_{\min}^{-1}\left\{d(a_{1} + a_{2})E[\mathcal{N}\{r-2\}]
    + da_{2}E[\mathcal{N}\{r-3\}]\right\}^{1/2}  \right)  \\
    &\quad\quad \le w_{\min}^{-1}\left\{d(a_{1} + 2a_{2})\right\}^{3/2},
\end{align}
which implies ${\rm{Var}}[u_{\boldsymbol{\mathcal{\check{A}}}_{i}}]=O(n^{1/2})$ as stated. The first inequality is from \eqref{eq:VUABound}; the second is due to summing over $i$; the third is by recognising that $\mathcal{N}\{r\}/n\le 1$ for all $r$ and simplifying. Note, it is very likely that $\mathcal{N}\{r\}<<n^{1/2}$ for large $r$. If this were verified then the approximation bound, $O(n^{1/2})$, could be considerably refined; however, the result as stated is sufficient for the current work. This concludes the proof of Lemma \ref{lem:VV}. 

\section{Proof of Lemma \ref{eq:Tr0}}
First it is shown that $\boldsymbol{D}$ is positive semi-definite. Let $\boldsymbol{P} = \boldsymbol{H} \boldsymbol{M}_{\xi}^{-1} \boldsymbol{H}^{T}$ denote the hat matrix which is well known to be positive semi-definite, where $\boldsymbol{H}$ is a $d\times p$ matrix with $i$th row equal to $\boldsymbol{\dot{\eta}}(\boldsymbol{x}_{i})$. Therefore, $\boldsymbol{D}$ can be expressed as $\boldsymbol{D} = \boldsymbol{R}\boldsymbol{P}\boldsymbol{R} = \boldsymbol{R}\boldsymbol{P}^{1/2}\boldsymbol{P}^{1/2}\boldsymbol{R}$,
where $\boldsymbol{R}= {\rm{diag}}(\boldsymbol{r})$. This implies $\boldsymbol{D}$ is positive semi-definite. 

Let $\boldsymbol{\overline{W}} = (\boldsymbol{W} - \boldsymbol{w}\boldsymbol{w}^{T})$; if $d=1$ then $tr(\boldsymbol{D}\boldsymbol{\overline{W}}) = 0$ and equality is achieved. If $d>1$ then $\boldsymbol{\overline{W}}$ is positive definite and $\lambda_{\min}(\boldsymbol{\overline{W}})>0$. Since $\boldsymbol{D}$ is positive semi-definite, \citet{Klei:Athan:TheD:1968} implies that
\begin{align} \label{eq:trBound}
    tr(\boldsymbol{D}\boldsymbol{\overline{W}}) &\ge \lambda_{\min}(\boldsymbol{\overline{W}})tr(\boldsymbol{D}) \ge 0
\end{align}
Therefore, if $d>1$ \eqref{eq:trBound} can only equal 0 only if $tr(\boldsymbol{D})=0$. It is useful to use the expanded form of $tr(\boldsymbol{D})$ written as  
\begin{align}
    tr(\boldsymbol{D}) &= \sum_{i=1}^{d}\sum_{k=1}^{d} \boldsymbol{c}^{T}\boldsymbol{M}_{\xi}^{-1} \boldsymbol{\dot{\eta}}(\boldsymbol{x}_{i}) \boldsymbol{\dot{\eta}}^{T}(\boldsymbol{x}_{i}) \boldsymbol{M}_{\xi}^{-1} \boldsymbol{\dot{\eta}}(\boldsymbol{x}_{k}) \boldsymbol{\dot{\eta}}^{T}(\boldsymbol{x}_{k})\boldsymbol{M}_{\xi}^{-1}\boldsymbol{c} \\
    &\ge \boldsymbol{c}^{T}\left[\sum_{i=1}^{d} w_{i}\boldsymbol{M}_{\xi}^{-1} \boldsymbol{\dot{\eta}}(\boldsymbol{x}_{i}) \boldsymbol{\dot{\eta}}^{T}(\boldsymbol{x}_{i}) \boldsymbol{M}_{\xi}^{-1} \sum_{k=1}^{d} w_{k}\boldsymbol{\dot{\eta}}(\boldsymbol{x}_{k}) \boldsymbol{\dot{\eta}}^{T}(\boldsymbol{x}_{k}) \boldsymbol{M}_{\xi}^{-1}\right]\boldsymbol{c} \\
    &= \boldsymbol{c}^{T}\left[\boldsymbol{M}_{\xi}^{-1}\sum_{i=1}^{d} w_{i} \boldsymbol{\dot{\eta}}(\boldsymbol{x}_{i}) \boldsymbol{\dot{\eta}}^{T}(\boldsymbol{x}_{i}) \boldsymbol{M}_{\xi}^{-1} \right]\boldsymbol{c} \\
    &= \boldsymbol{c}^{T}\boldsymbol{M}_{\xi}^{-1}\boldsymbol{c} \\
    &>0
\end{align}
for all non-zero $\boldsymbol{c}$ per the condition that inverse of $\boldsymbol{M}_{\xi}$ exists. This concludes the proof.

\end{supplement}



\bibliographystyle{imsart-nameyear.bst}
\bibliography{bibtex_entries}

\end{document}